\documentclass[letterpaper,twocolumn,pra,aps,superscriptaddress,amsmath,amssymb,floatfix,article]{revtex4-1}

\usepackage[utf8]{inputenc}
\usepackage{array}
\usepackage{wrapfig}
\usepackage{tabularx}

\usepackage{graphicx}
\usepackage{dcolumn}
\usepackage{bm}
\usepackage{xcolor}
\usepackage{multirow}
\usepackage{graphicx}
\usepackage{subfigure}
\usepackage{float}

\usepackage{tikz}
\usepackage{ulem}
\usepackage{CJK}

\begin{document}
\begin{CJK*}{UTF8}{gbsn} 

\preprint{APS/123-QED} 
\title{Line intensities of CO near 1560 nm measured with absorption and dispersion spectroscopy}

\author{Q. Huang (黄琪)$^\ddag$}
\affiliation{
    State Key Laboratory of Molecular Reaction Dynamics, Department of Chemical Physics, University of Science and Technology of China, Hefei 230026, China}
\author{Y. Tan (谈艳)$^\ddag$}%
\affiliation{
    Hefei National Research Center of Physical Sciences at the Microscale, University of Science and Technology of China, Hefei 230026, China}
\author{R.-H. Yin (尹睿恒)}
\affiliation{
    Department of Physics, University of Science and Technology of China, Hefei 230026, China}
\author{Z.-L. Nie (聂中梁)}%
\affiliation{
    State Key Laboratory of Molecular Reaction Dynamics, Department of Chemical Physics, University of Science and Technology of China, Hefei 230026, China}
\author{J. Wang (王进)}%
    \email{jinwang@ustc.edu.cn}
\affiliation{
    Hefei National Research Center of Physical Sciences at the Microscale, University of Science and Technology of China, Hefei 230026, China}
\author{S.-M. Hu (胡水明)}%
    \email{smhu@ustc.edu.cn}
\affiliation{
    State Key Laboratory of Molecular Reaction Dynamics, Department of Chemical Physics, University of Science and Technology of China, Hefei 230026, China}
\affiliation{
    Hefei National Laboratory, University of Science and Technology of China, Hefei 230088, China}
\thanks{Authors equally contributed to this work.}%

\date{\today}

\begin{abstract}

High-precision line intensities are of great value in various applications, such as greenhouse gas metrology, planetary atmospheric analysis, and trace gas detection. 
Here we report simultaneous measurements of cavity-enhanced absorption and dispersion spectroscopy of the prototype molecule $^{12}$C$^{16}$O using the same optical resonant cavity. Nine lines were measured in the R branch of the $v=3-0$ band. The absorption and dispersion spectra were fitted separately with speed-dependent Voigt profiles, and the line intensities obtained by the two methods agree within the experimental uncertainty of about 1\textperthousand. 
The results demonstrate the feasibility of SI-traceable molecular density measurements based on laser spectroscopy.

\end{abstract}

\maketitle

\section{Introduction}


Greenhouse gases (GHGs) such as carbon dioxide, methane, carbon monoxide, and nitrous oxide are measured by global satellite remote sensing and terrestrial platforms to monitor the natural background and identify emitting sources. Most existing quantitative measurement methods and instruments require calibration with Standard Reference Materials (SRMs)~\cite{Rhoderick2018}. Although optical methods based on molecular absorption spectroscopy are widely used nowadays, conventional gas metrology relies on the gravimetric method. However, discrepancies arising from absorption on the internal surface of the cylinder and the gas fractionation during standard mixtures preparation often exceed the compatibility target (0.1 $\mu$mol/mol for CO$_2$) recommended by the World Meteorological Organization (WMO)~\cite{Aoki2022}. 

In recent years, it has become increasingly attractive to realize greenhouse gas metrology traceable to the International System of Units (SI), which aims to establish a direct link between molecular absorption spectroscopy and SI units~\cite{Henningsen2000JMS-CO2, Casa2007JCP-CO2, Padilla2007IEEE, Fleisher2019PRL}. These systems facilitate spectroscopic instruments with unprecedented accuracy, enabling direct calibration of remote sensing satellites and ground-based platforms and improving the accuracy of in situ measurements~\cite{Polyansky2015PRL-CO2, Fleisher2019PRL, Reed2020Optica, Bielska2022PRL-CO}.
However, accurate measurements of CO$_2$ line intensities are often complicated by the adsorption and desorption of CO$_2$ on the inner surfaces of measurement cavities~\cite{Long2020GRL-CO2, Birk2021JQSRT-CO2}.
A few CO$_2$ lines in the near-infrared and infrared regions up to better than 3\textperthousand\, have been reported with a joint experimental and theoretical study providing rotation-vibration line intensities with the required accuracy. 
It is still very difficult to reach the WMO criteria of 0.1 $\mu$mol/mol of CO$_2$ in the air, which corresponds to a fractional uncertainty of 0.2\textperthousand\, in the line intensities.
Meanwhile, CO is a more favorable candidate to validate calculations and experimental methods. Its reduced susceptibility to adsorption and desorption effects allows for an accurate determination of the pressure.
As a simple diatomic molecule, CO has simpler energy levels with less overlap of adjacent lines and isotopologues in its infrared spectrum, and the transition moments could be calculated more accurately. These characteristics greatly facilitate measurements with a higher accuracy.

The semi-empirical dipole moment of carbon monoxide and line lists for all its isotopologues have been reconstructed and checked by simultaneous nonlinear least squares fitting (NLLSF) of the selected experimental intensities for $^{12}$C$^{16}$O and the ab initio permanent dipole moment~\cite{Meshkov2022JQSRT-CO}. 
Some problems have been found with the previous CO line list in HITRAN calculated by Li \textit{et al.}\cite{Li2015ApJSS-CO, HITRAN2020}.
Line positions in the near-infrared second overtone band (3-0) of $^{12}$C$^{16}$O have been determined by cavity-enhanced spectroscopy with $10^{-10}$ accuracy~\cite{Cygan2016JCP, Cygan2019OE-3method} or even higher~\cite{Wang2017JCP, Wang2021JQSRT-CO}. 
Cygan \textit{et al.}~\cite{Cygan2019OE-3method} also reported the intensities of the R(23), R(24), and R(28) lines with relative uncertainties at the parts per thousand level. 
The intensities of the P(26) to P(28) lines were reported with similar accuracy by Reed \textit{et al.} using frequency-stabilized cavity ring-down spectroscopy~\cite{Reed2017-CO}. 
However, there are many controversies between different data sources, especially for intensities at the sub-percent level. 
Recently, accurate intensities of lines in the (3-0) band of $^{12}$C$^{16}$O have been reported independently from three laboratories, including the cavity ring-down spectroscopy (CRDS) at the National Institute of Standards and Technology (NIST), cavity-mode dispersion spectroscopy (CMDS) at the Nicolaus Copernicus University (NCU), and the Fourier-transform spectroscopy (FTS) at Physikalische Technische Bundesanstalt (PTB). The agreement between the experiment and the theoretical calculation from University College London (UCL) reaches the level of 1\textperthousand~\cite{Bielska2022PRL-CO}, shedding light on the SI-traceable determination of the CO concentration in gas samples.

Here we report the measurement of CO line intensities by both cavity-enhanced absorption spectroscopy and cavity-enhanced dispersion spectroscopy with one high-finesse cavity. Metrological-grade temperature and pressure calibration techniques were applied to the measurement of several lines in the (3-0) band of the main isotopologue $^{12}$C$^{16}$O. Line intensities with uncertainties at the 1\textperthousand\, level were derived by fitting the spectra. 
In the study of Bielska \textit{et al.}~\cite{Bielska2022PRL-CO}, the CMDS measurements performed at NCU used pure CO gas samples with pressures up to 13~kPa, and the CRDS measurements conducted in NIST used mixing CO-N$_2$ gas samples with pressures up to 26~kPa. It is mandatory to use sophisticated Hartmann-Tran profile (HTP)~\cite{Ngo2013JQSRT-HTP, Tran2014JQSRT-Err} including both velocity- and phase-changing collision effects in high pressures. Note that the line profile model contributes the largest uncertainty (1\textperthousand) in the results given by Bielska \textit{et al.}~\cite{Bielska2022PRL-CO}. Since considerably lower sample pressures were applied in this work, allowing us to use the relatively simple speed-dependent Voigt Profile (SDV) instead of HTP. We expect that using a simpler line profile model involving fewer effective parameters can avoid unphysical results due to correlations among multiple parameters in the fitting.
Compared to the results presented by Bielska \textit{et al.}, we also extended to cover a few more transitions with $J''_{max} = 32$. 
Moreover, both CRDS and CMDS measurements conducted in the same experimental setup also allow for the investigation of {type-B} uncertainties in both methods.

\section{Experimental \label{sec:exp}}

\begin{figure}[htp]
\centering{}\includegraphics[width=\columnwidth]{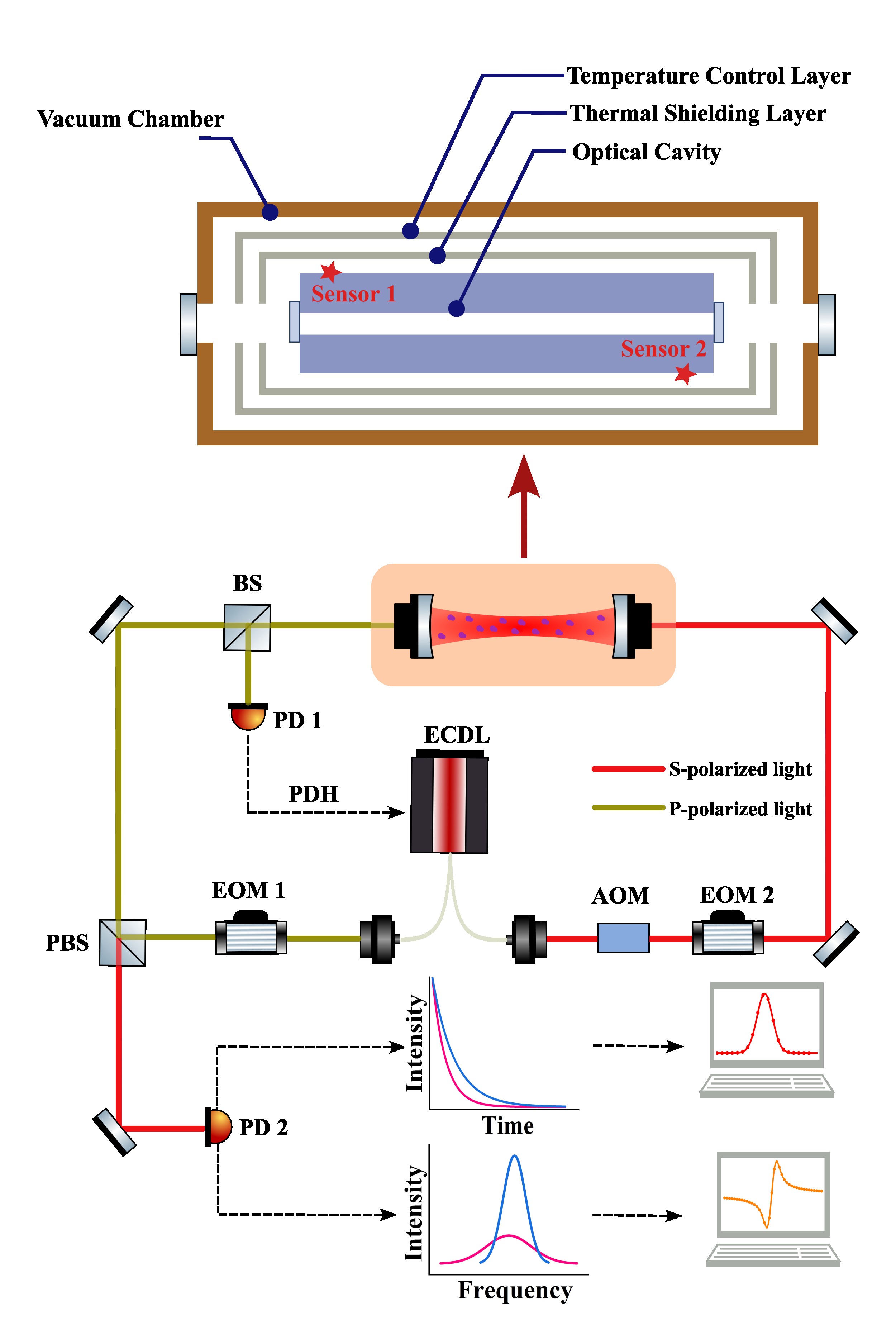}
\caption{Configuration of the experimental setup for cavity ring-down spectroscopy (CRDS) and cavity mode dispersion spectroscopy (CMDS). Abbreviations: AOM, acousto-optic modulator; BS, beam splitter; ECDL, external-cavity diode laser; EOM, electro-optic modulator; PBS: polarized beam splitter; PD, photodiodes.  
}
\label{fig:setup}
\end{figure} 

Cavity ring-down spectroscopy (CRDS) and cavity mode dispersion spectroscopy (CMDS) were applied to measure the CO lines near 1560~nm using the same high-finesse cavity. The experimental setup is similar to that presented in Refs.~\cite{cheng2015, Wang2017JCP, Tao2018}.
A schematic configuration of the experimental setup is shown in Fig.~\ref{fig:setup}.
The optical cavity was composed of a pair of high-reflective (HR) mirrors ($R = 99.997\%$ at 1.5-1.7 $\mu$m), separated by a distance of 118 cm, corresponding to a free-spectral range of 126.9 MHz, and a mode width of about 0.6 kHz.
An aluminum tube was used to support the mirror mounts for the HR mirrors. Two glass capsule thermal sensors, SPRTs (5686-B, Fluke), were placed on either side of the tube as shown in Fig.~\ref{fig:setup}. 
The optical cavity was placed inside a stainless steel vacuum chamber, and the chamber was attached with heating pads for coase temperature control. 
Two aluminum alloy cylinders were placed between the stainless steel chamber and the optical cavity. 
A heating wire was attached to the outer aluminum cylinder and controlled by a locking servo to stabilize the temperature.
The inner aluminum cylinder was used as a heat shield.
In this way, the temperature of the optical cavity can be precisely controlled, and the temperature of the gas sample at equilibrium should be identical to the temperature read by the two SPRTs.

An external cavity diode laser (ECDL, Toptica DL Pro) was used as the probe laser source.
A beam from the laser, referred to as the ``locking beam'',  was introduced into the optical cavity and the frequency was stabilized using the Pound-Drever-Hall (PDH) technique.
Another laser beam, referred to as the ``probing beam'', was frequency-shifted via an acoustic-optical modulator (AOM) and a fiber electro-optical modulator (EOM) and then introduced into the cavity from the opposite side of the cavity. Frequency scanning was accomplished by switching the radiofrequency applied to the EOM. When the sum of the frequencies applied to the AOM and EOM matched a longitudinal mode of the cavity, the probe beam passed through the cavity and was detected by a photodiode. All radio frequencies were synchronized to a GPS-disciplined rubidium clock (SRS FS725).

\begin{figure}[htp]
\centering
\includegraphics[width=0.95\columnwidth]{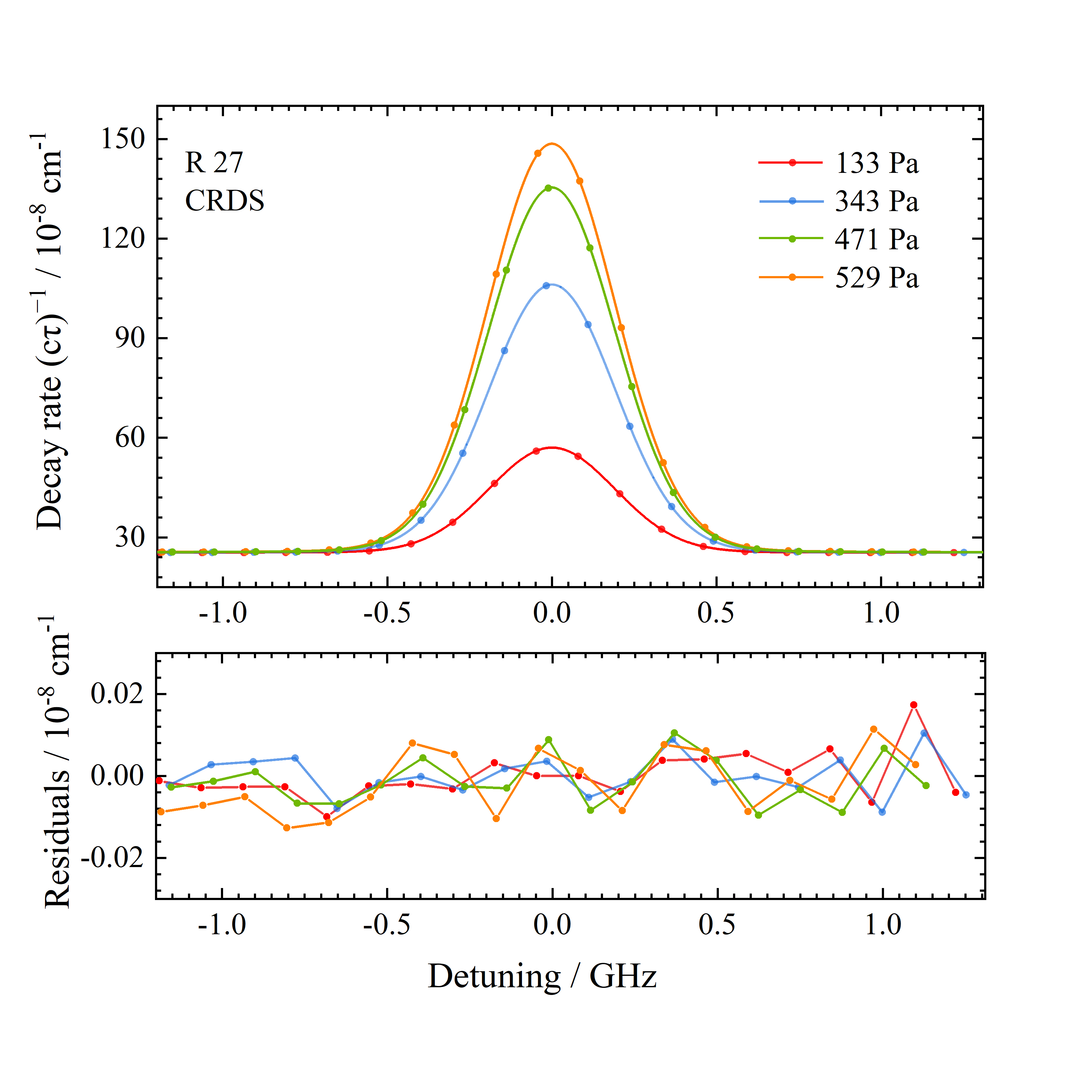}
\caption{
Upper panel: CRDS spectra of the R(27) line of $^{12}$C$^{16}$O (single scan) measured at 299 K, with a pure CO sample gas of 194, 343, 471, and 529~Pa. Experimental data are shown as dots, and the solid lines indicate simulated spectra. Bottom panel: Fitting residuals. Speed-dependent Voigt (SDV) profiles were used in the fit.
}
\label{fig:R27CRDS}
\end{figure}
\begin{figure}[htbp]
\centering
\includegraphics[width=.95\columnwidth]{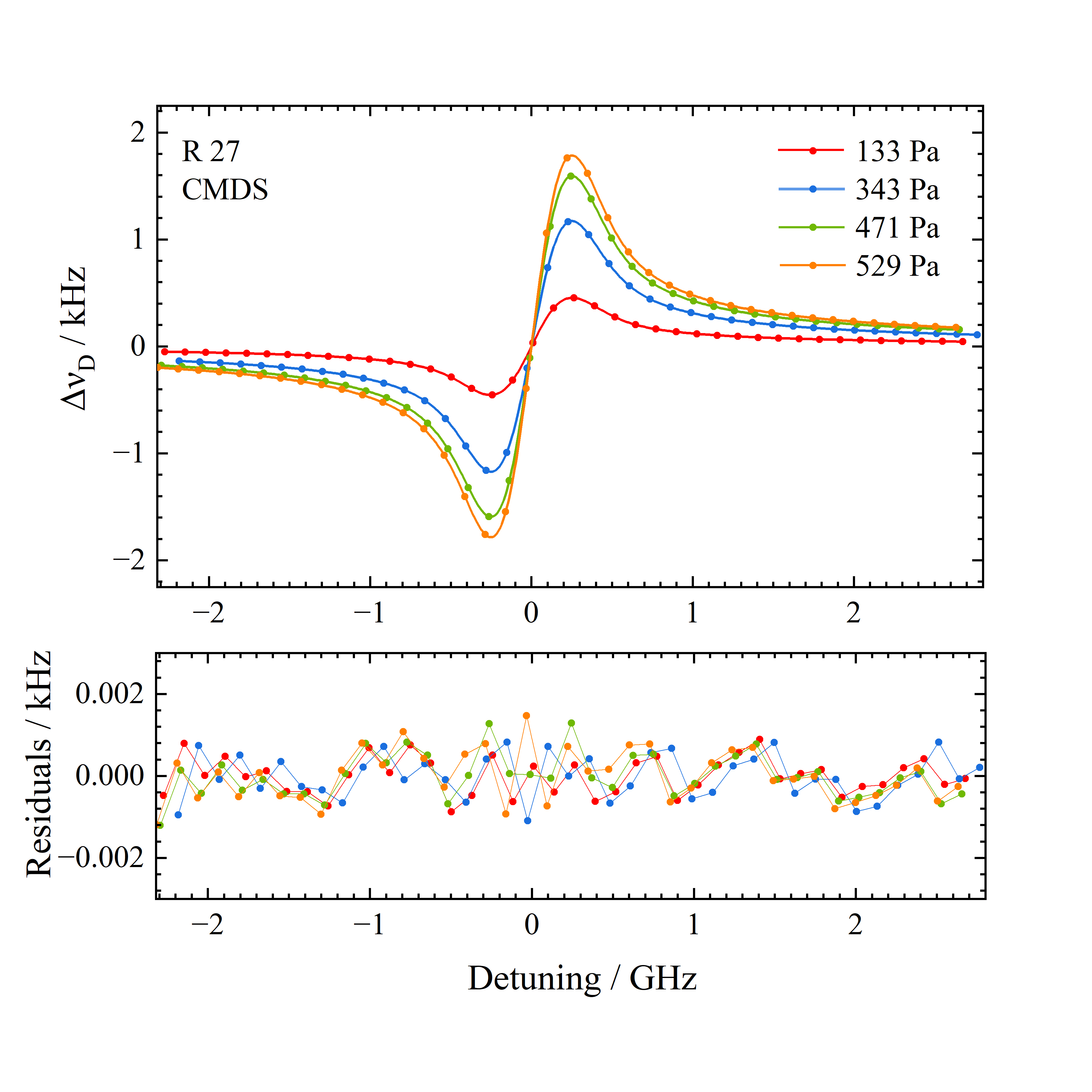}
\caption{
Upper panel: CMDS spectra of the R(27) line of $^{12}$C$^{16}$O (single scan) measured at 299 K, with a pure CO sample gas of 133, 343, 471, and 529~Pa. Experimental data are shown as dots, and the solid lines indicate simulated spectra. Bottom panel: Fitting residuals. Speed-dependent Voigt (SDV) profiles were used in the fit. 
}
\label{fig:R27CMDS}
\end{figure}

In the CRDS mode, the ring-down event was initiated by switching off the RF signal applied to the fiber EOM, the ring-down signal was recorded with a digitizer (NI PXI 5922), and the curve was fitted with an exponential decay function to determine the decay time $\tau$. 
In the CMDS mode, transmission spectra of tens of cavity modes were recorded to derive precise mode positions. Changes in the mode positions due to the molecular absorption line were measured, giving the sample-induced dispersion spectra.
Typically, one CRDS scan took about 4 seconds and one CMDS scan took about 11 seconds. 
Different sample pressures up to 1.3 kPa were used in the measurements.
Fig.~\ref{fig:R27CRDS} and Fig.~\ref{fig:R27CMDS} show CRDS and CMDS spectra of the R(27) line recorded at different pressures, respectively. 

\section{Data Analysis \label{sec:analysis}}
CRDS measures the absorption spectrum, while CMDS measures the dispersion spectrum. The line shape models for both spectra are:
\begin{eqnarray}
    \alpha(\nu_m) &=& A\times \operatorname{Re}\left[\varphi\left(\nu_m-\nu_{c}\right) \right] \label{eq:abs}\\
    \frac{\Delta\nu }{\nu_{m}} &=& A \times \frac{\operatorname{Im}\left[\varphi\left(\nu_m-\nu_{c}\right)\right]} {2 nk_{0}} \label{eq:disp}
\end{eqnarray}
where $A$ is the integrated intensity of the absorption line,
$\nu_{m}$ is the frequency of the mode with index of $m$ equal to the laser frequency plus the microwave frequency $(f_{AOM}+f_{EOM})$; 
$n$ is the frequency-independent refractive index of the gas, and $k_{0}$ is the wave vector of the transition frequency.
Within a range of over 5~GHz around the absorption line, we can ignore the dispersion due to the cavity mirrors.~\cite{Rutkowski2017OE-cavdisp}
The frequency shift of the cavity mode due to the dispersion induced by the molecular absorption line is $\Delta\nu$.
The line shapes of the absorption and dispersion spectra are the real and imaginary parts of the normalized line shape function~\cite{Libbrecht2006AJP-refractive, Cygan2015OE-1DCMDS} $\varphi\left(\nu_m-\nu_{c}\right)$, respectively.
The absorption line strength (in cm/molecule) can be derived from the experimental peak area $A$ at temperature $T$ and pressure $P$:
\begin{eqnarray} 
    S(T) &=& A \times \frac{k_BT}{P}, \label{eq:S} \\
    \frac{S(T_0)}{S(T)} &=& e^{\frac{hcE''}{k_B}(\frac{1}{T}-\frac{1}{T_0})}
        [\frac{1-e^{-hc\nu_c/k_BT_0}}{1-e^{-hc\nu_c/k_BT}}]\frac{Q(T)}{Q(T_0)},
        \label{eq:ST0}
\end{eqnarray}
where $k_B$ is the Boltzmann constant, $h$ is the Planck constant, $c$ is the speed of light, $E''$ is the energy of the lower state of the transition, and $Q(T)$ is the partition sum~\cite{Gamache2021JQSRT-Q} at temperature $T$.
The line strength $S(T)$ could be converted~\cite{HITRAN2020} to the HITRAN standard temperature value at $T_0=296$~K according to Eq.~(\ref{eq:ST0}).

The recorded spectra were fitted with the speed-dependent Voigt (SDV) profile.
The SDV line shape parameters include the center of the line $\nu_c$, the peak area $A$, the Doppler half width $\Gamma_D$, the Lorentzian half width averaged by velocity $\Gamma_0$, the speed-dependent relaxation rate $\Gamma_2$, the velocity-averaged line shift $\Delta_0$ and the speed-dependent line shift $\Delta_2$.
The SDV profile is equivalent to a simplified version of the Hartmann-Tran profile (HTP)~\cite{Ngo2013JQSRT-HTP, Tran2014JQSRT-Err}, with both the velocity-changing collision parameter $\nu_{VC}$ and the correlation parameter $\eta$ fixed at zero.
In this work, only four parameters $\nu_0$, $A$, $\Gamma_0$, and $\Gamma_2$ were relaxed in the fit. 
The Doppler half-width was fixed to the calculated value at the experimental temperature.
The speed-dependent line shift parameter $\Delta_2$ was set to zero as all measurements were carried out under rather low pressures.
The line shift parameter $\Delta_0$ was also excluded from the fit since its contribution was included in the floating parameter $\nu_c$ and all spectra recorded under different pressures were fitted independently. 
As an example, the CRDS spectra of the R(27) line together with the fit residuals are shown in Fig.~\ref{fig:R27CRDS}.

\begin{figure}[htbp]
\centering
\includegraphics[width=.95\columnwidth]{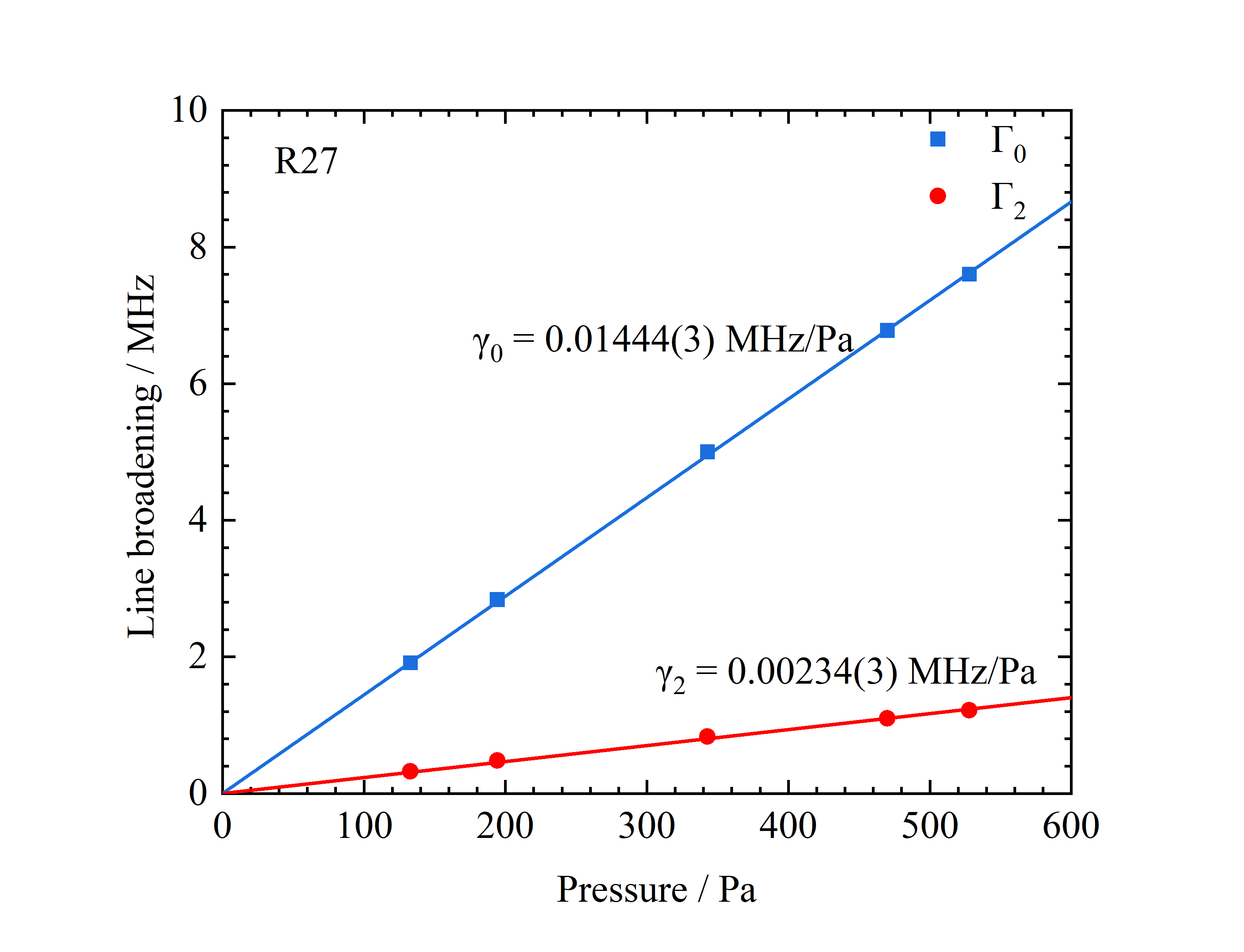}
\caption{
Line broadening parameters $\Gamma_0$ and $\Gamma_2$ obtained from fitting CRDS spectra of the R(27) line recorded under different pressures. Speed-dependent Voigt (SDV) profiles were used in the fit. Coefficients  $\gamma_0$ and $\gamma_2$ were determined by a linear fit of the data. }
\label{fig:gamma_CRDS}
\end{figure}

Fig.~\ref{fig:gamma_CRDS} shows the line broadening parameters $\Gamma_0$ and $\Gamma_2$ obtained by fitting CRDS spectra of the R(27) line recorded under different pressures. We can see that they follow an excellent linear dependence on the pressures~\cite{Ngo2013JQSRT-HTP, Tran2013JQSRT-HTP, Tran2014JQSRT-Err}. The broadening coefficients $\gamma_0$ and $\gamma_2$ were determined by a linear fit of the data, and the relative uncertainties were about 2\textperthousand~ and 13\textperthousand, respectively.


The signal-to-noise ratio of the CMDS spectra was about an order of magnitude worse than that of the CRDS spectra. Non-physical parameters could be obtained by relaxing all parameters in a non-Voigt profile to fit the CMDS spectra. According to Eq.~\ref{eq:abs} and Eq.~\ref{eq:disp}, both absorption and dispersion spectra should have the same profile function. When fitting the CMDS spectra, we fixed the $\gamma_0$ and $\gamma_2$ coefficients to those obtained from fitting the CRDS spectra.
Fig.~\ref{fig:R27CMDS} shows the CMDS spectra of the R(27) line fitted with SDV profiles. The fit residuals are at the noise level, indicating an excellent consistency between the absorption and dispersion spectra.

The HTP model~\cite{Ngo2013JQSRT-HTP, Tran2014JQSRT-Err} is considered complete describing the absorption spectrum under arbitrary pressures, but quite a few parameters in this model have not been accurately determined elsewhere experimentally or theoretically. 
Since low sample pressures were used in this work, the simpler SDV model is sufficient to fit the spectra observed here.
However, we can use the HTP model to estimate the potential systematic uncertainty in line intensity due to the uncertainty in the line profile model.
We implemented the HTP model and floated the speed-dependent line shift parameter $\Delta_2$ and the velocity-changing coefficient $\nu_{VC}$ in the fit, but the resulting change in the peak area $A$ was within \textcolor{black}{an} uncertainty of 0.3\textperthousand. 
Therefore, we conservatively give an uncertainty of 0.3\textperthousand~ to account for the possible {type-B} uncertainty due to the line profile model.

\section{Uncertainty budget}
In this section, we analyze the uncertainties in the line intensities obtained from the absorption and dispersion spectra. The R(27) line is used as an example.
The uncertainties from different sources are discussed below, and the uncertainty budget is given in Table~\ref{tab:budget}.
Uncertainties from different sources are categorized into type-A and type-B uncertainties, and the overall uncertainty was derived as $u_{tot} = (\sum u_i^2)^{1/2}$.

\begin{table}
\caption{\label{tab:budget} 
    Error budget ($k=1$) of the line intensity of R(27) (unit: permille, \textperthousand) }
\begin{tabular}{ccc}
\hline
Source & $\mathrm{u_{r}}$(CRDS) & $\mathrm{u_{r}}$(CMDS) \\
 & \textperthousand & \textperthousand \\
 \hline
\multicolumn{3}{c}{\textbf{Type A}} \\
Statistical & 0.05 & 0.4 \\
\hline
 \multicolumn{3}{c}{\textbf{Type B}} \\
Sample gas purity   & 0.1  & 0.1 \\
Isotopic abundance of $^{12}$C$^{16}$O & 0.2 & 0.2 \\
Pressure            & 0.5& 0.5\\
Temperature         & 0.1  & 0.1 \\
Line shape          & 0.3  & 0.3 \\
Others              & $<0.3$ & $<0.3$ \\
\hline
Total & 0.7& 0.8\\
\hline
\end{tabular}
\end{table}


\subsection{Statistical} 

\begin{figure*}[htp]
    \centering
    \includegraphics[width=0.8\textwidth]{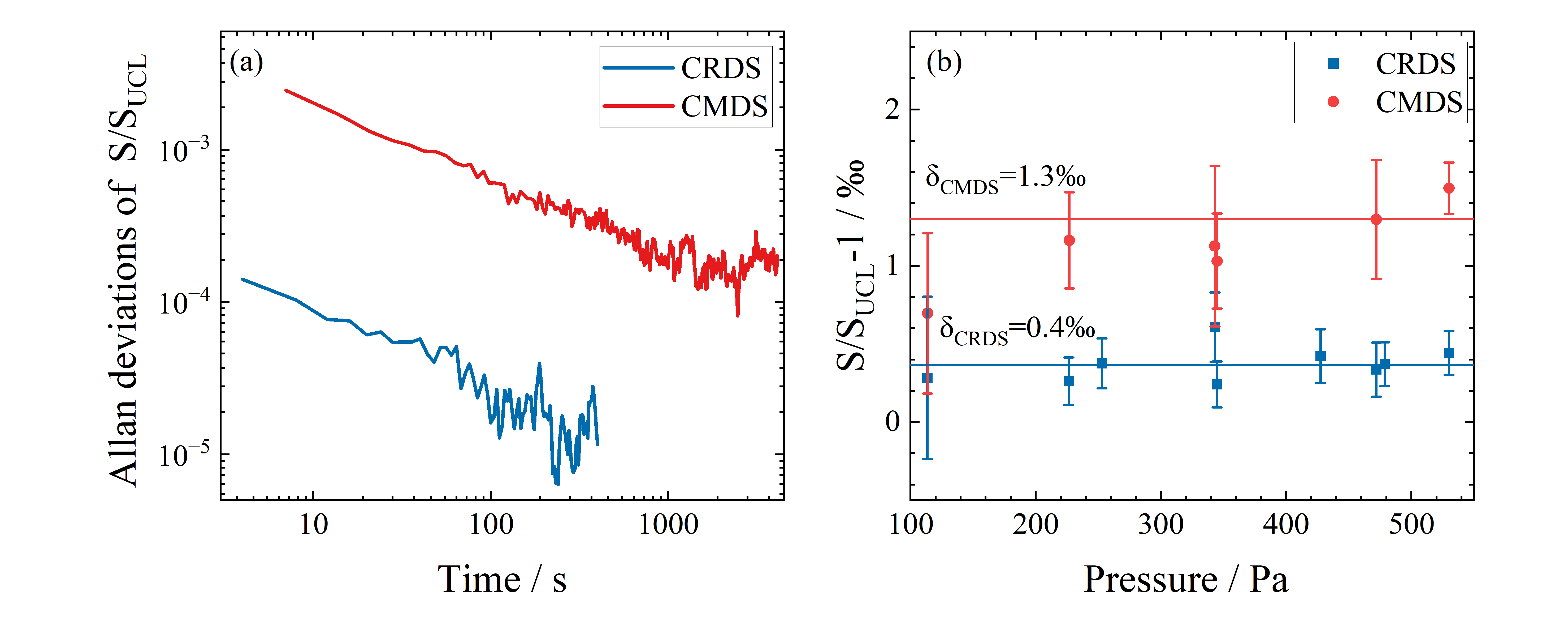}
    \caption{(a) {Allan deviations of the intensities of the R(27) line intensities obtained by CRDS (blue curve) and CMDS (red curve). All experimental values were divided by the UCL calculated value of $1.5049\times 10^{-25}$~cm/molecule.}
    (b) Comparison of the CRDS and CMDS line intensities obtained at different pressures. Data are shown as the deviations from the UCL calculated values. 
    {Note that the error bars only represent the type-A uncertainties. Type-B uncertainties are not included here.}
    \label{fig:Allan} }
\end{figure*}

For each spectral line, several thousand spectra were recorded at each pressure. 
Fig.~\ref{fig:Allan}(a) illustrates the fractional Allan deviation of the line intensities derived from fitting the spectra of the R(27) line. It can be seen that the Allan deviation, which represents the {type-A} uncertainty of average value, decreases to 0.02\textperthousand\, after averaging hundreds of CRDS spectra or 0.2\textperthousand\, after averaging several thousand CMDS spectra.
The signal-to-noise ratio for a single CMDS spectrum is about an order of magnitude lower than for CRDS. As a result, we typically acquired CMDS spectra ten times more frequently than CRDS scans.
Fig.~\ref{fig:Allan}(b) depicts the line intensities obtained from all recorded spectra at different pressures. For better illustration, deviations from the UCL calculated values are shown in the figure. Notably, the CRDS values show reasonable agreement with the CMDS values. We found a systematic deviation between the values obtained from the two methods: the averaged CRDS intensity of R(27) is less than the CMDS value by 0.9\textperthousand, which is considerably larger than the statistical uncertainty. A similar trend was also found for other lines studied in this work.
In principle, the CMDS method could be immune to many sources of systematic errors~\cite{Cygan2019OE-3method} since it measures only the frequencies, and frequency is the quantity that could be measured most precisely. 
Cygan \textit{et al.} also reported~\cite{Cygan2016MST} systematic deviations between CRDS and CMDS in the line intensity measurement. They observed a pressure dependence of the CRDS-measured line intensity. 
Here we did not observe such dependence in our CRDS measurement within the range of 100-500~Pa, as shown in Fig.~\ref{fig:Allan}(b).
The systematic deviation between our CRDS and CMDS measurements seems larger than the statistical uncertainty but comparable to the overall uncertainty (see the following subsections). Further investigations would be feasible if we could reduce the uncertainties from other sources.


\subsection{Gas Purity}

The sample gas used in this experiment is pure CO gas (AirLiquid co., stated purity $ > 99.99$\%). A liquid-N$_2$ cold trap was applied to remove possible contaminants in the sample before use. 

\begin{figure}[htp]
\centering{} \includegraphics[width=0.95\columnwidth]{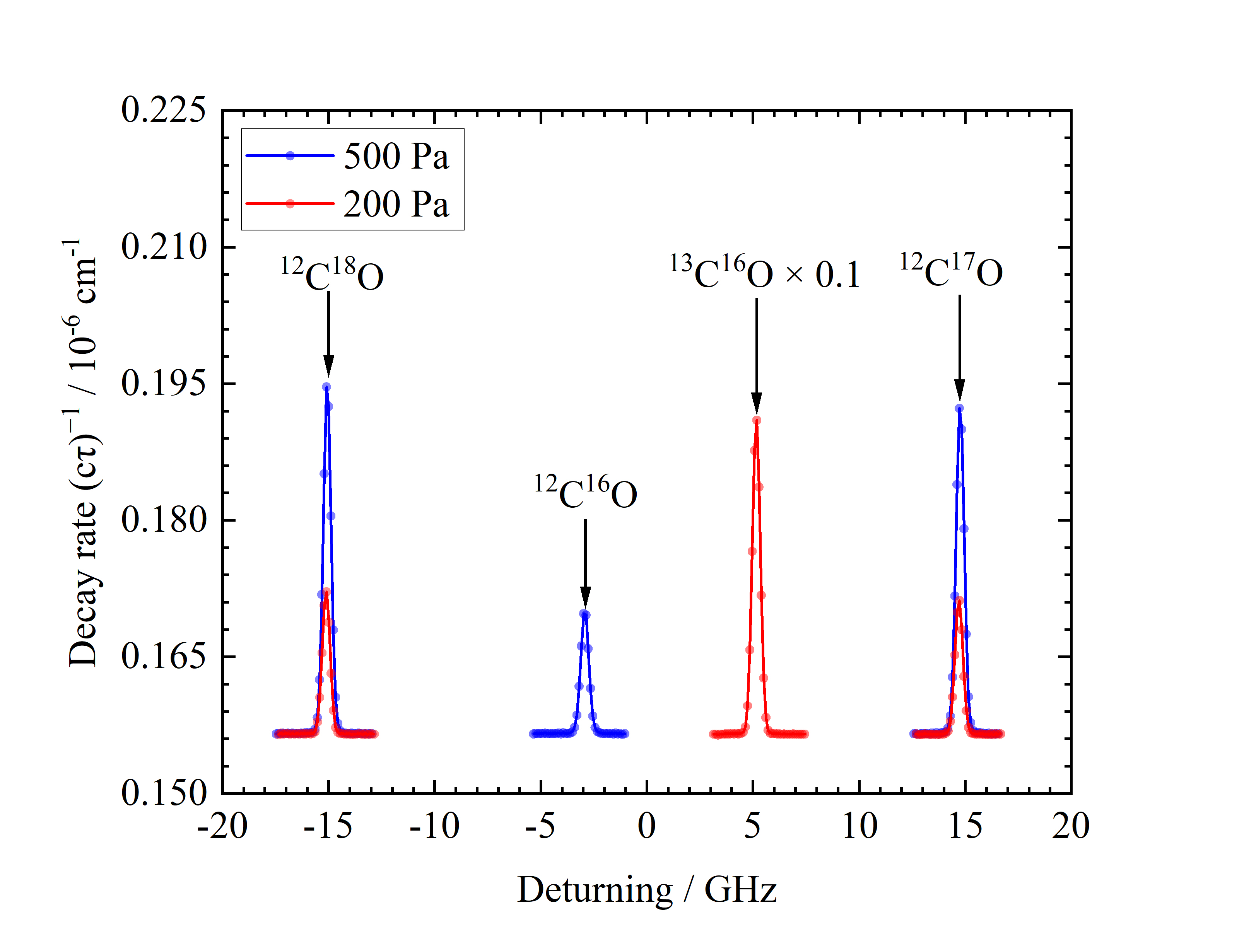}
\caption{CRDS spectra of the CO sample gas near 6255 cm$^{-1}$ recorded at 200 Pa (red) and 500 Pa (blue). Abundances of three minor isotopologues relative to the main isotopologue were determined from the spectra. Note the spectrum of the $^{13}$C$^{16}$O isotopologue has been multiplied by a factor of 0.1.
}
\label{fig:CO-iso}
\end{figure} 

The HITRAN recommended abundances of 6 isotopologues of CO are given in Table~\ref{tab:isotope}. We used the CRDS absorption spectra to determine the abundances of three major isotopologues after the main isotopologue $^{12}$C$^{16}$O in our gas sample: $^{13}$C$^{16}$O, $^{12}$C$^{18}$O, and $^{12}$C$^{17}$O. Four lines in the range of 6254.8 - 6256.0 cm$^{-1}$ were measured, corresponding to $^{12}$C$^{16}$O, $^{12}$C$^{17}$O, $^{13}$C$^{16}$O, and $^{12}$C$^{18}$O, respectively. 
Two less abundant species in the sample, $^{13}$C$^{18}$O and $^{13}$C$^{17}$O, are neglected since their abundances are below 0.1~\textperthousand. We estimated the abundances of $^{12}$C$^{17}$O, $^{13}$C$^{16}$O, and $^{12}$C$^{18}$O from the absorptions obtained from the experimental spectra and the line intensities given in the HITRAN database. The uncertainties of the HITRAN line intensities are less than 2\%, so we obtained the abundances of these three minor isotopologues with a 2\% uncertainty, as given in Table~\ref{tab:isotope}. We can see that they reasonably agree with the natural abundance recommended by HITRAN. The abundance of the main isotopologue in our sample was determined to be 98.699(22)\%, which is slightly higher than the value of 98.6544\% recommended by HITRAN. 

\begin{table}[htp]
    \centering
    \caption{Isotope abundance (value in parenthesis is the $1\sigma$ uncertainty in the last quoted digit) }
    \begin{tabular}{ l|l|l } 
    \hline
         Isotope  & this work  & HITRAN\\ 
    \hline
         $^{13}$C$^{16}$O & 0.01063(21)  & 0.0110836\\ 
         $^{12}$C$^{18}$O & 0.00200(4) & 0.00197822 \\ 
         $^{12}$C$^{17}$O & 0.000379(8)& 0.000367867\\ 
         $^{13}$C$^{18}$O &  -         & 0.000022225 \\ 
         $^{13}$C$^{17}$O &  -         & 0.00000413292 \\ 
    \hline
         $^{12}$C$^{16}$O & 0.98699(22)  & 0.986544\\ 
    \hline
    \end{tabular}
    \label{tab:isotope}
\end{table}

\subsection{Pressure Measurement}

\begin{figure}[htbp]
\centering 
    \includegraphics[width=0.95\columnwidth]{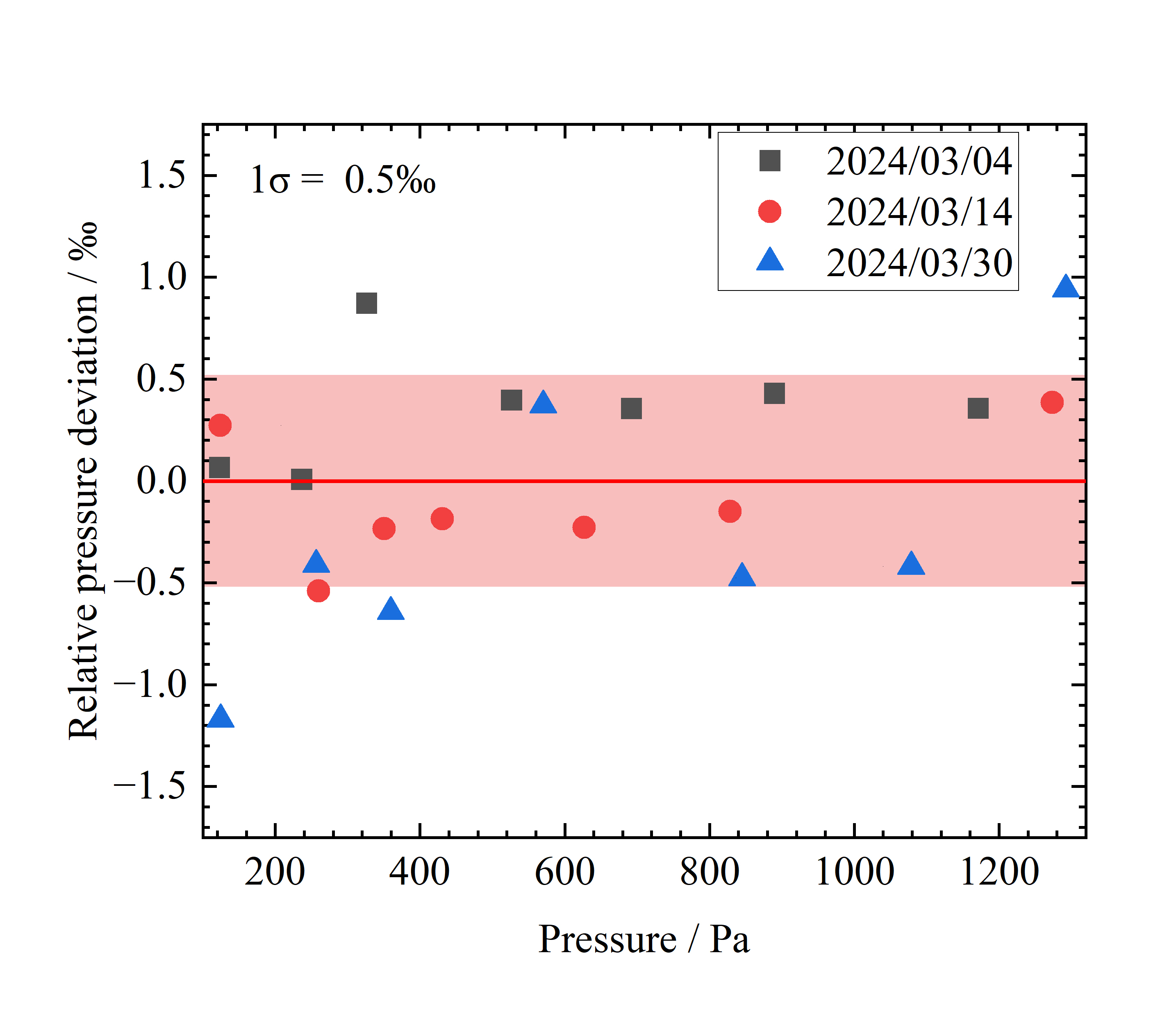}
    \caption{Deviations ($P_{mano}/P_{opt}-1$) of the calibrated Leybold manometer from the optical pressure gauge. The working gas was pure argon. Data points measured on different dates were shown in different colors. 
        \label{fig:Leybold}
        }
\end{figure}

A capacitance manometer (Leybold CTR101N) with a stated uncertainty of 1.2\textperthousand\, was used in the experiment. 
The accuracy of the manometer was verified by comparing it to an optical pressure gauge several times. 
The optical pressure gauge was based on a Fabry-P\'{e}rot interferometer made of ultra-low-expansion (ULE) glass, and it was calibrated by a piston gauge from the National Institute of Metrology (Beijing) using the working gas of pure argon (AirLiquid Co., purity $>99.999\%$).~\cite{Nie2024} After calibration, the optical pressure gauge has an accuracy of 0.012~\textperthousand~($k=1$) when measuring pure argon. 
Fig.~\ref{fig:Leybold} shows the differences between the Leybold manometer and the optical pressure gauge when measuring pure argon gases. The pressure range is similar to the CO sample gas pressures used in this work. We can see that the manometer readings agree excellently with the optical gauge. 
As shown in the figure, the $1\sigma$ deviation of the pressure readings is 0.5\textperthousand, and we use this value as the uncertainty of the pressure measurement in this work.
Before each measurement, the sample cell was pumped by a turbo pump to below 1~mPa before refilling the gas sample. The pressure drift in the cell was measured to be below the uncertainty of the pressure gauge.

\subsection{Temperature Measurement}

\begin{figure}[htbp]
    \centering
    \includegraphics[width=0.95\columnwidth]{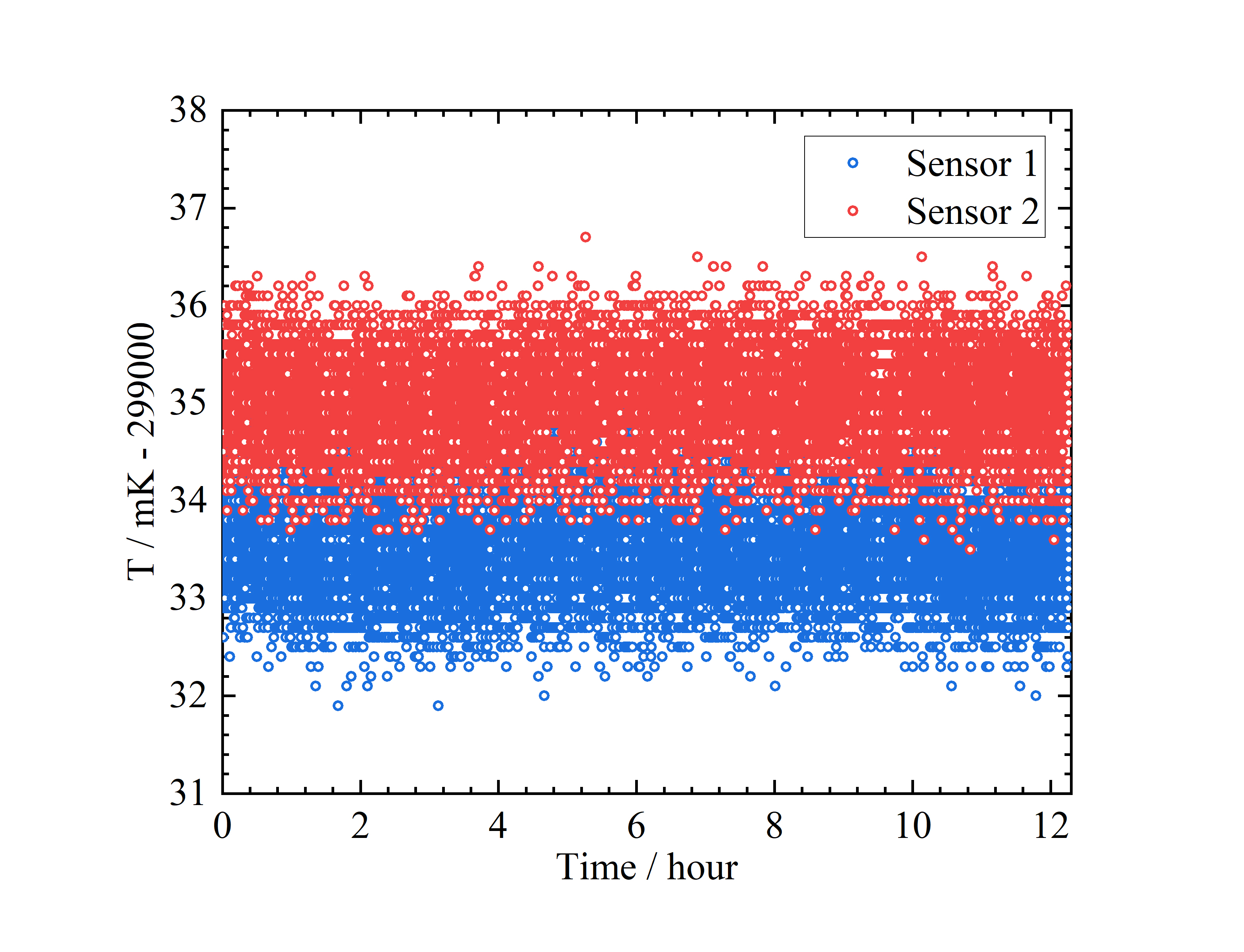}
    \caption{Temperature drift of the ring-down cavity measured by two SPRTs sensors. The sample gas pressure was about 410~Pa.} 
    \label{fig:T2}
\end{figure}

The temperature drift during the measurement was monitored by two SPRTs (5686-B, Fluke) sensors placed at both ends of the ring-down cavity. 
Both the sensors and readout (MKT50, Anton Parr) were calibrated at the National Institute of Metrology (Beijing) and the uncertainties were less than 5~mK.
After filling the gas sample, we waited about one hour before the measurement to confirm that the gas temperature had been at equilibrium with the sample cell. 
Fig.~\ref{fig:T2} illustrates the temperature drift recorded by the two sensors over 12 hours. The results show that the fluctuation was only about 1~mK and the difference between the two sensors was less than 3~mK, indicating excellent temperature stability of the sample cavity. The contribution of the temperature measurement to the line intensity, including the correction for the temperature-dependent population of molecules, was estimated to be less than 0.1\textperthousand for R(27).

\subsection{Others}

Nonlinearity in data acquisition could introduce deviations in the retrieved line intensity from the spectrum, especially for CRDS measurements. 
The nonlinearity of the data acquisition (DAQ) card could affect the accuracy of CRDS measurements~\cite{Fleisher2019PRL}. The one we used in this work is the NI PXI 5922 (4 MSa/s, 20-bit) card, which is considered to be a highly linear digitizer~\cite{Overney2011IEEE} and has been used for accurate CRDS measurements by the NIST group~\cite{Fleisher2019PRL, Bielska2022PRL-CO}. 
We also tried another digitizer in the CRDS measurement: NI 9223 (1 MSa/s, 16-bit). 
The CRDS integrated absorbance of the R(27) line at 230~Pa obtained with these two DAQ cards shows a deviation of 0.02(30)\textperthousand. We repeated these comparisons several times and confirmed that the deviation due to the DAQ card should remain below 0.2\textperthousand.


High intra-cavity laser power induces saturation effects, potentially leading to deviations in the results. In this study, we minimized the input laser power, and the intra-cavity laser power was estimated to be less than 0.2~W. We deliberately recorded spectra under different laser powers, and the discrepancies among the results remained within 0.2\textperthousand.

\section{Summary and Discussion}

\begin{figure}[htp]
    \centering
    \includegraphics[width=0.95\columnwidth]{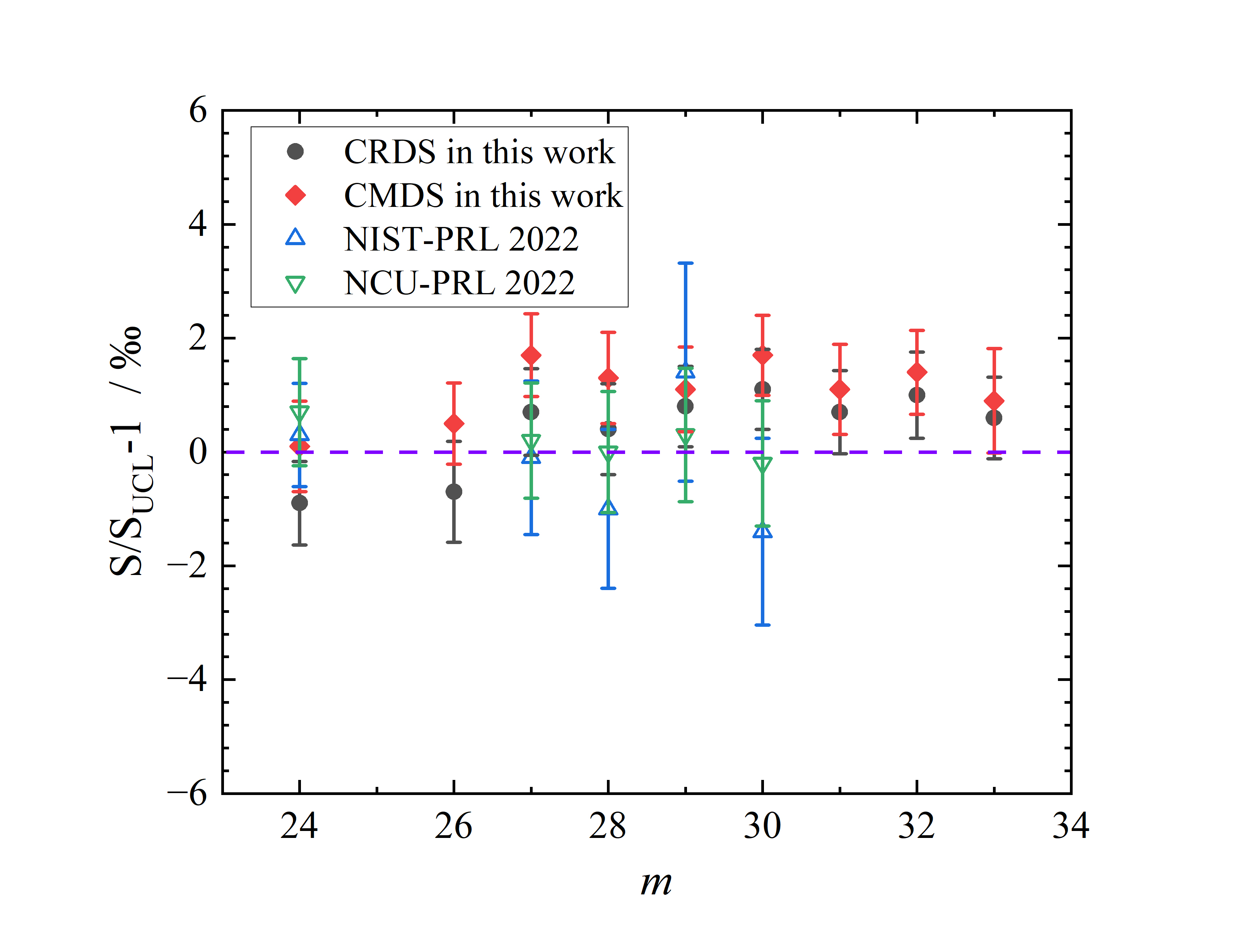}
    \caption{Comparison of line intensities obtained in this work and literature. The horizontal axis is $m=J+1$ for transitions in the R branch. All values were normalized to the calculated values given by the UCL group.
    \label{fig:result}
    }
\end{figure}

In total, we measured 9 lines of $^{12}$C$^{16}$O in the (3-0) band within the R branch for rotation numbers up to 32. 
The results are given in Table~\ref{tab:result}. 
Line positions in the table are from our previous Lamb-dip measurements~\cite{Wang2021JQSRT-CO}.
The line intensities obtained from this work range from $1\times 10^{-26}$~cm/molecule to $8\times 10^{-25}$~cm/molecule, and the fractional uncertainties are in the range of 0.7-0.8~\textperthousand. 
The UCL calculated line intensities are also given in the table.
The CMDS line intensities obtained in this work are larger than the UCL values by {1.1\textperthousand\,} in average, and also larger than our CRDS values by 0.7\textperthousand\, in average.
Deviations between the UCL calculated results and experimental values from NIST and NCU~\cite{Bielska2022PRL-CO} are also illustrated in Fig.~\ref{fig:result}. All the values agree within a combined uncertainty below 2\textperthousand.

The main {type-B} uncertainty in this work comes from the pressure measurement (0.5\textperthousand).
It also prevents us from extending the high precision measurements to more lines in this band {(particularly those $J''\leq 20$ lines)} at this stage. Transitions with lower rotation numbers are interesting to compare with FTS measurements~\cite{Bielska2022PRL-CO}. However, these lines are stronger and need to be measured at lower pressures for pure samples, while it is a challenge to accurately determine the gas pressure below 100~Pa.
Gas absorption/desorption from the cavity walls would become more serious at such low pressures.
Alternatively, these lines could be measured with mixed gases, as has been done by the NIST group~\cite{Polyansky2015PRL-CO2, Bielska2022PRL-CO}, provided that the CO concentrations in the samples are known accurately.
{A better way is to transport the bottles of mixed CO/N$_2$ gases among different laboratories and compare the experimental results obtained with the same samples, and the possibility is under planning.}
Transitions with higher rotation numbers are weaker and need to be measured at higher pressures to achieve a sufficient signal-to-noise ratio. 
The current pressure gauge used in this work can only reach 1.3~kPa, and a new gauge covering higher pressures will be used in subsequent studies.
Since the population of molecules at high rotational levels is more sensitive to temperature, the contribution to the uncertainty from the temperature measurement will increase.
The uncertainty from the line profile model also needs to be considered at high pressures, and it is necessary to use accurate (sophisticated) line profiles such as the Hartmann-Tran profile (HTP)~\cite{Ngo2013JQSRT-HTP}.
For the measurements performed in this work at pressures below 1~kPa, we found that the relatively simple speed-dependent Voigt profile is sufficient for the analysis with an accuracy of 1\textperthousand.~\cite{Wojtewicz2013JQSRT-CO-Profile}

The establishment of SI-traceable determination of molecular densities based on absorption spectroscopy mandates accurate intensity measurements at the $10^{-4}$ level to meet the requirements of various applications. Cavity-enhanced spectroscopy methods are well suited for such measurements, where both a high signal-to-noise ratio and wide dynamic range are essential. Many sources of uncertainty considered in this work, such as isotopic abundance, pressure and temperature measurements, and line profile models, should be analyzed to improve the measurement accuracy. Cavity-enhanced mode dispersion spectroscopy (CMDS) measures the frequency shifts, eliminating many systematic errors in data acquisition, and is considered more promising for quantitative measurements. However, as shown in this work, the sensitivity of CMDS is still worse than that of cavity ring-down spectroscopy (CRDS) by an order of magnitude. The comparison between CMDS, CRDS, and other spectroscopic methods with precision at the 1\textperthousand\, level or better would be very useful to investigate the systematic uncertainties of different methods, which is necessary for an SI-traceable measurement towards the $10^{-4}$ accuracy.

\begin{table*}
\caption{\label{tab:result}
 CO line intensities obtained in this work, with 100\% $^{12}$C$^{16}$O isotopic abundance. 
 }
\begin{ruledtabular}
\begin{tabular}{ccc|cc|cc}
 & Position$^*$ & $I^{calc.}_{\textrm{UCL}}$ 
    & \multicolumn{2}{c|}{CRDS, \textit{absorption}}
    & \multicolumn{2}{c}{CMDS, \textit{dispersion}} \\
 & cm$^{-1}$ & cm/molecule 
    & $I^{exp.}_{\textrm{CRDS}}$, cm/molecule & $\delta^\dag$, \textperthousand 
    & $I^{exp.}_{\textrm{CMDS}}$, cm/molecule & $\delta^\dag$, \textperthousand \\
\hline
 R(23)& 6410.879543 & 8.1663E-25 & 8.1592(60)E-25& -0.9& 8.1673(65)E-25&0.1\\
 R(25)& 6413.122496 & 3.6481E-25 & 3.6455(32)E-25& -0.7& 3.6501(26)E-25&0.5\\
 R(26)& 6414.080821& 2.3663E-25 & 2.3679(18)E-25&  0.7& 2.3703(17)E-25&1.7\\
 R(27)& 6414.930187& 1.5049E-25 & 1.5055(11)E-25&  0.4& 1.5069(12)E-25&1.3\\
 R(28)& 6415.670442& 9.3839E-26 & 9.3917(66)E-26&  0.8& 9.3946(70)E-26&1.1\\
 R(29)& 6416.301445& 5.7381E-26 & 5.7443(40)E-26&  1.1& 5.7481(41)E-26&1.7\\
 R(30)& 6416.823044 & 3.4411E-26 & 3.4436(25)E-26&  0.7& 3.4449(27)E-26&1.1\\
 R(31)& 6417.235084 & 2.0240E-26 & 2.0261(15)E-26&  1.0& 2.0268(15)E-26&1.4\\
 R(32)& 6417.537471 & 1.1677E-26 & 1.1684(8)E-26& 0.6& 1.1688(11)E-26&0.9\\
\end{tabular}
\end{ruledtabular}
\flushleft
$^*$ Positions from Lamb-dip measurements~\cite{Wang2021JQSRT-CO}. \\
$^\dag$ $\delta = I^{exp.}/I^{calc.}_{\textrm{UCL}} -1 $;
\end{table*}


\begin{acknowledgments}
This work was jointly supported by the National Natural Science Foundation of China (Grants No. 12393822, 12393825), and the Ministry of Science and Technology of China (Grant Nos. 2021ZD0303102, 2022YFF0606500).
\end{acknowledgments} 

\end{CJK*} 
 

\begin{thebibliography}{32}%
\makeatletter
\providecommand \@ifxundefined [1]{%
 \@ifx{#1\undefined}
}%
\providecommand \@ifnum [1]{%
 \ifnum #1\expandafter \@firstoftwo
 \else \expandafter \@secondoftwo
 \fi
}%
\providecommand \@ifx [1]{%
 \ifx #1\expandafter \@firstoftwo
 \else \expandafter \@secondoftwo
 \fi
}%
\providecommand \natexlab [1]{#1}%
\providecommand \enquote  [1]{``#1''}%
\providecommand \bibnamefont  [1]{#1}%
\providecommand \bibfnamefont [1]{#1}%
\providecommand \citenamefont [1]{#1}%
\providecommand \href@noop [0]{\@secondoftwo}%
\providecommand \href [0]{\begingroup \@sanitize@url \@href}%
\providecommand \@href[1]{\@@startlink{#1}\@@href}%
\providecommand \@@href[1]{\endgroup#1\@@endlink}%
\providecommand \@sanitize@url [0]{\catcode `\\12\catcode `\$12\catcode
  `\&12\catcode `\#12\catcode `\^12\catcode `\_12\catcode `\%12\relax}%
\providecommand \@@startlink[1]{}%
\providecommand \@@endlink[0]{}%
\providecommand \url  [0]{\begingroup\@sanitize@url \@url }%
\providecommand \@url [1]{\endgroup\@href {#1}{\urlprefix }}%
\providecommand \urlprefix  [0]{URL }%
\providecommand \Eprint [0]{\href }%
\providecommand \doibase [0]{http://dx.doi.org/}%
\providecommand \selectlanguage [0]{\@gobble}%
\providecommand \bibinfo  [0]{\@secondoftwo}%
\providecommand \bibfield  [0]{\@secondoftwo}%
\providecommand \translation [1]{[#1]}%
\providecommand \BibitemOpen [0]{}%
\providecommand \bibitemStop [0]{}%
\providecommand \bibitemNoStop [0]{.\EOS\space}%
\providecommand \EOS [0]{\spacefactor3000\relax}%
\providecommand \BibitemShut  [1]{\csname bibitem#1\endcsname}%
\let\auto@bib@innerbib\@empty
\bibitem [{\citenamefont {Rhoderick}\ \emph {et~al.}(2018)\citenamefont
  {Rhoderick}, \citenamefont {Kelley}, \citenamefont {Miller}, \citenamefont
  {Norris}, \citenamefont {Carney}, \citenamefont {Gameson}, \citenamefont
  {Cecelski}, \citenamefont {Harris}, \citenamefont {Goodman}, \citenamefont
  {Srivastava},\ and\ \citenamefont {Hodges}}]{Rhoderick2018}%
  \BibitemOpen
  \bibfield  {author} {\bibinfo {author} {\bibfnamefont {G.~C.}\ \bibnamefont
  {Rhoderick}}, \bibinfo {author} {\bibfnamefont {M.~E.}\ \bibnamefont
  {Kelley}}, \bibinfo {author} {\bibfnamefont {W.~R.}\ \bibnamefont {Miller}},
  \bibinfo {author} {\bibfnamefont {J.~E.}\ \bibnamefont {Norris}}, \bibinfo
  {author} {\bibfnamefont {J.}~\bibnamefont {Carney}}, \bibinfo {author}
  {\bibfnamefont {L.}~\bibnamefont {Gameson}}, \bibinfo {author} {\bibfnamefont
  {C.~E.}\ \bibnamefont {Cecelski}}, \bibinfo {author} {\bibfnamefont {K.~J.}\
  \bibnamefont {Harris}}, \bibinfo {author} {\bibfnamefont {C.~A.}\
  \bibnamefont {Goodman}}, \bibinfo {author} {\bibfnamefont {A.}~\bibnamefont
  {Srivastava}}, \ and\ \bibinfo {author} {\bibfnamefont {J.~T.}\ \bibnamefont
  {Hodges}},\ }\href {\doibase 10.1021/acs.analchem.7b05310} {\bibfield
  {journal} {\bibinfo  {journal} {Anal. Chem.}\ }\textbf {\bibinfo {volume}
  {90}},\ \bibinfo {pages} {4711} (\bibinfo {year} {2018})}\BibitemShut
  {NoStop}%
\bibitem [{\citenamefont {Aoki}\ \emph {et~al.}(2022)\citenamefont {Aoki},
  \citenamefont {Ishidoya}, \citenamefont {Murayama},\ and\ \citenamefont
  {Matsumoto}}]{Aoki2022}%
  \BibitemOpen
  \bibfield  {author} {\bibinfo {author} {\bibfnamefont {N.}~\bibnamefont
  {Aoki}}, \bibinfo {author} {\bibfnamefont {S.}~\bibnamefont {Ishidoya}},
  \bibinfo {author} {\bibfnamefont {S.}~\bibnamefont {Murayama}}, \ and\
  \bibinfo {author} {\bibfnamefont {N.}~\bibnamefont {Matsumoto}},\ }\href
  {\doibase 10.5194/amt-15-5969-2022} {\bibfield  {journal} {\bibinfo
  {journal} {Atmos. Meas. Technol.}\ }\textbf {\bibinfo {volume} {15}},\
  \bibinfo {pages} {5969} (\bibinfo {year} {2022})}\BibitemShut {NoStop}%
\bibitem [{\citenamefont {Henningsen}\ and\ \citenamefont
  {Simonsen}(2000)}]{Henningsen2000JMS-CO2}%
  \BibitemOpen
  \bibfield  {author} {\bibinfo {author} {\bibfnamefont {J.}~\bibnamefont
  {Henningsen}}\ and\ \bibinfo {author} {\bibfnamefont {H.}~\bibnamefont
  {Simonsen}},\ }\href@noop {} {\bibfield  {journal} {\bibinfo  {journal} {J.
  Mol. Spectrosc.}\ }\textbf {\bibinfo {volume} {203}},\ \bibinfo {pages} {16}
  (\bibinfo {year} {2000})}\BibitemShut {NoStop}%
\bibitem [{\citenamefont {Casa}\ \emph {et~al.}(2007)\citenamefont {Casa},
  \citenamefont {Parretta}, \citenamefont {Castrillo}, \citenamefont {Wehr},\
  and\ \citenamefont {Gianfrani}}]{Casa2007JCP-CO2}%
  \BibitemOpen
  \bibfield  {author} {\bibinfo {author} {\bibfnamefont {G.}~\bibnamefont
  {Casa}}, \bibinfo {author} {\bibfnamefont {D.~A.}\ \bibnamefont {Parretta}},
  \bibinfo {author} {\bibfnamefont {A.}~\bibnamefont {Castrillo}}, \bibinfo
  {author} {\bibfnamefont {R.}~\bibnamefont {Wehr}}, \ and\ \bibinfo {author}
  {\bibfnamefont {L.}~\bibnamefont {Gianfrani}},\ }\href@noop {} {\bibfield
  {journal} {\bibinfo  {journal} {J. Chem. Phys.}\ }\textbf {\bibinfo {volume}
  {127}} (\bibinfo {year} {2007})}\BibitemShut {NoStop}%
\bibitem [{\citenamefont {Padilla-Viquez}\ \emph {et~al.}(2007)\citenamefont
  {Padilla-Viquez}, \citenamefont {Koelliker-Delgado}, \citenamefont {Werhahn},
  \citenamefont {Jousten},\ and\ \citenamefont {Schiel}}]{Padilla2007IEEE}%
  \BibitemOpen
  \bibfield  {author} {\bibinfo {author} {\bibfnamefont {G.}~\bibnamefont
  {Padilla-Viquez}}, \bibinfo {author} {\bibfnamefont {J.}~\bibnamefont
  {Koelliker-Delgado}}, \bibinfo {author} {\bibfnamefont {O.}~\bibnamefont
  {Werhahn}}, \bibinfo {author} {\bibfnamefont {K.}~\bibnamefont {Jousten}}, \
  and\ \bibinfo {author} {\bibfnamefont {D.}~\bibnamefont {Schiel}},\ }\href
  {\doibase 10.1109/TIM.2007.891160} {\bibfield  {journal} {\bibinfo  {journal}
  {IEEE Trans. Instrum. Meas.}\ }\textbf {\bibinfo {volume} {56}},\ \bibinfo
  {pages} {529} (\bibinfo {year} {2007})}\BibitemShut {NoStop}%
\bibitem [{\citenamefont {Fleisher}\ \emph {et~al.}(2019)\citenamefont
  {Fleisher}, \citenamefont {Adkins}, \citenamefont {Reed}, \citenamefont {Yi},
  \citenamefont {Long}, \citenamefont {Fleurbaey},\ and\ \citenamefont
  {Hodges}}]{Fleisher2019PRL}%
  \BibitemOpen
  \bibfield  {author} {\bibinfo {author} {\bibfnamefont {A.~J.}\ \bibnamefont
  {Fleisher}}, \bibinfo {author} {\bibfnamefont {E.~M.}\ \bibnamefont
  {Adkins}}, \bibinfo {author} {\bibfnamefont {Z.~D.}\ \bibnamefont {Reed}},
  \bibinfo {author} {\bibfnamefont {H.}~\bibnamefont {Yi}}, \bibinfo {author}
  {\bibfnamefont {D.~A.}\ \bibnamefont {Long}}, \bibinfo {author}
  {\bibfnamefont {H.~M.}\ \bibnamefont {Fleurbaey}}, \ and\ \bibinfo {author}
  {\bibfnamefont {J.~T.}\ \bibnamefont {Hodges}},\ }\href {\doibase
  10.1103/physrevlett.123.043001} {\bibfield  {journal} {\bibinfo  {journal}
  {Phys. Rev. Lett.}\ }\textbf {\bibinfo {volume} {123}},\ \bibinfo {pages} {1}
  (\bibinfo {year} {2019})}\BibitemShut {NoStop}%
\bibitem [{\citenamefont {Polyansky}\ \emph {et~al.}(2015)\citenamefont
  {Polyansky}, \citenamefont {Bielska}, \citenamefont {Ghysels}, \citenamefont
  {Lodi}, \citenamefont {Zobov}, \citenamefont {Hodges},\ and\ \citenamefont
  {Tennyson}}]{Polyansky2015PRL-CO2}%
  \BibitemOpen
  \bibfield  {author} {\bibinfo {author} {\bibfnamefont {O.~L.}\ \bibnamefont
  {Polyansky}}, \bibinfo {author} {\bibfnamefont {K.}~\bibnamefont {Bielska}},
  \bibinfo {author} {\bibfnamefont {M.}~\bibnamefont {Ghysels}}, \bibinfo
  {author} {\bibfnamefont {L.}~\bibnamefont {Lodi}}, \bibinfo {author}
  {\bibfnamefont {N.~F.}\ \bibnamefont {Zobov}}, \bibinfo {author}
  {\bibfnamefont {J.~T.}\ \bibnamefont {Hodges}}, \ and\ \bibinfo {author}
  {\bibfnamefont {J.}~\bibnamefont {Tennyson}},\ }\href@noop {} {\bibfield
  {journal} {\bibinfo  {journal} {Phys. Rev. Lett.}\ }\textbf {\bibinfo
  {volume} {114}} (\bibinfo {year} {2015})}\BibitemShut {NoStop}%
\bibitem [{\citenamefont {Reed}\ \emph {et~al.}(2020)\citenamefont {Reed},
  \citenamefont {Long}, \citenamefont {Fleurbaey},\ and\ \citenamefont
  {Hodges}}]{Reed2020Optica}%
  \BibitemOpen
  \bibfield  {author} {\bibinfo {author} {\bibfnamefont {Z.~D.}\ \bibnamefont
  {Reed}}, \bibinfo {author} {\bibfnamefont {D.~A.}\ \bibnamefont {Long}},
  \bibinfo {author} {\bibfnamefont {H.}~\bibnamefont {Fleurbaey}}, \ and\
  \bibinfo {author} {\bibfnamefont {J.~T.}\ \bibnamefont {Hodges}},\ }\href
  {\doibase 10.1364/OPTICA.395943} {\bibfield  {journal} {\bibinfo  {journal}
  {Optica}\ }\textbf {\bibinfo {volume} {7}},\ \bibinfo {pages} {1209}
  (\bibinfo {year} {2020})}\BibitemShut {NoStop}%
\bibitem [{\citenamefont {Bielska}\ \emph {et~al.}(2022)\citenamefont
  {Bielska}, \citenamefont {Kyuberis}, \citenamefont {Reed}, \citenamefont
  {Li}, \citenamefont {Cygan}, \citenamefont {Ciury\'{l}o}, \citenamefont
  {Adkins}, \citenamefont {Lodi}, \citenamefont {Zobov}, \citenamefont {Ebert},
  \citenamefont {Lisak}, \citenamefont {Hodges}, \citenamefont {Tennyson},\
  and\ \citenamefont {Polyansky}}]{Bielska2022PRL-CO}%
  \BibitemOpen
  \bibfield  {author} {\bibinfo {author} {\bibfnamefont {K.}~\bibnamefont
  {Bielska}}, \bibinfo {author} {\bibfnamefont {A.~A.}\ \bibnamefont
  {Kyuberis}}, \bibinfo {author} {\bibfnamefont {Z.~D.}\ \bibnamefont {Reed}},
  \bibinfo {author} {\bibfnamefont {G.}~\bibnamefont {Li}}, \bibinfo {author}
  {\bibfnamefont {A.}~\bibnamefont {Cygan}}, \bibinfo {author} {\bibfnamefont
  {R.}~\bibnamefont {Ciury\'{l}o}}, \bibinfo {author} {\bibfnamefont {E.~M.}\
  \bibnamefont {Adkins}}, \bibinfo {author} {\bibfnamefont {L.}~\bibnamefont
  {Lodi}}, \bibinfo {author} {\bibfnamefont {N.~F.}\ \bibnamefont {Zobov}},
  \bibinfo {author} {\bibfnamefont {V.}~\bibnamefont {Ebert}}, \bibinfo
  {author} {\bibfnamefont {D.}~\bibnamefont {Lisak}}, \bibinfo {author}
  {\bibfnamefont {J.~T.}\ \bibnamefont {Hodges}}, \bibinfo {author}
  {\bibfnamefont {J.}~\bibnamefont {Tennyson}}, \ and\ \bibinfo {author}
  {\bibfnamefont {O.~L.}\ \bibnamefont {Polyansky}},\ }\href {\doibase
  10.1103/PhysRevLett.129.043002} {\bibfield  {journal} {\bibinfo  {journal}
  {Phys. Rev. Lett.}\ }\textbf {\bibinfo {volume} {129}},\ \bibinfo {pages}
  {043002} (\bibinfo {year} {2022})}\BibitemShut {NoStop}%
\bibitem [{\citenamefont {Long}\ \emph {et~al.}(2020)\citenamefont {Long},
  \citenamefont {Reed}, \citenamefont {Fleisher}, \citenamefont {Mendonca},
  \citenamefont {Roche},\ and\ \citenamefont {Hodges}}]{Long2020GRL-CO2}%
  \BibitemOpen
  \bibfield  {author} {\bibinfo {author} {\bibfnamefont {D.~A.}\ \bibnamefont
  {Long}}, \bibinfo {author} {\bibfnamefont {Z.}~\bibnamefont {Reed}}, \bibinfo
  {author} {\bibfnamefont {A.~J.}\ \bibnamefont {Fleisher}}, \bibinfo {author}
  {\bibfnamefont {J.}~\bibnamefont {Mendonca}}, \bibinfo {author}
  {\bibfnamefont {S.}~\bibnamefont {Roche}}, \ and\ \bibinfo {author}
  {\bibfnamefont {J.~T.}\ \bibnamefont {Hodges}},\ }\href@noop {} {\bibfield
  {journal} {\bibinfo  {journal} {Geo. Res. Lett.}\ }\textbf {\bibinfo {volume}
  {47}},\ \bibinfo {pages} {e2019GL086344} (\bibinfo {year}
  {2020})}\BibitemShut {NoStop}%
\bibitem [{\citenamefont {Birk}\ \emph {et~al.}(2021)\citenamefont {Birk},
  \citenamefont {R{\"o}ske},\ and\ \citenamefont {Wagner}}]{Birk2021JQSRT-CO2}%
  \BibitemOpen
  \bibfield  {author} {\bibinfo {author} {\bibfnamefont {M.}~\bibnamefont
  {Birk}}, \bibinfo {author} {\bibfnamefont {C.}~\bibnamefont {R{\"o}ske}}, \
  and\ \bibinfo {author} {\bibfnamefont {G.}~\bibnamefont {Wagner}},\
  }\href@noop {} {\bibfield  {journal} {\bibinfo  {journal} {J. Quant.
  Spectrosc. Radiat. Transf.}\ }\textbf {\bibinfo {volume} {272}},\ \bibinfo
  {pages} {107791} (\bibinfo {year} {2021})}\BibitemShut {NoStop}%
\bibitem [{\citenamefont {Meshkov}\ \emph {et~al.}(2022)\citenamefont
  {Meshkov}, \citenamefont {Ermilov}, \citenamefont {Stolyarov}, \citenamefont
  {Medvedev}, \citenamefont {Ushakov},\ and\ \citenamefont
  {Gordon}}]{Meshkov2022JQSRT-CO}%
  \BibitemOpen
  \bibfield  {author} {\bibinfo {author} {\bibfnamefont {V.~V.}\ \bibnamefont
  {Meshkov}}, \bibinfo {author} {\bibfnamefont {A.~Y.}\ \bibnamefont
  {Ermilov}}, \bibinfo {author} {\bibfnamefont {A.~V.}\ \bibnamefont
  {Stolyarov}}, \bibinfo {author} {\bibfnamefont {E.~S.}\ \bibnamefont
  {Medvedev}}, \bibinfo {author} {\bibfnamefont {V.~G.}\ \bibnamefont
  {Ushakov}}, \ and\ \bibinfo {author} {\bibfnamefont {I.~E.}\ \bibnamefont
  {Gordon}},\ }\href@noop {} {\bibfield  {journal} {\bibinfo  {journal} {J.
  Quant. Spectrosc. Radiat. Transf.}\ }\textbf {\bibinfo {volume} {280}},\
  \bibinfo {pages} {108090} (\bibinfo {year} {2022})}\BibitemShut {NoStop}%
\bibitem [{\citenamefont {Li}\ \emph {et~al.}(2015)\citenamefont {Li},
  \citenamefont {Gordon}, \citenamefont {Rothman}, \citenamefont {Tan},
  \citenamefont {Hu}, \citenamefont {Kassi}, \citenamefont {Campargue},\ and\
  \citenamefont {Medvedev}}]{Li2015ApJSS-CO}%
  \BibitemOpen
  \bibfield  {author} {\bibinfo {author} {\bibfnamefont {G.}~\bibnamefont
  {Li}}, \bibinfo {author} {\bibfnamefont {I.~E.}\ \bibnamefont {Gordon}},
  \bibinfo {author} {\bibfnamefont {L.~S.}\ \bibnamefont {Rothman}}, \bibinfo
  {author} {\bibfnamefont {Y.}~\bibnamefont {Tan}}, \bibinfo {author}
  {\bibfnamefont {S.-M.}\ \bibnamefont {Hu}}, \bibinfo {author} {\bibfnamefont
  {S.}~\bibnamefont {Kassi}}, \bibinfo {author} {\bibfnamefont
  {A.}~\bibnamefont {Campargue}}, \ and\ \bibinfo {author} {\bibfnamefont
  {E.~S.}\ \bibnamefont {Medvedev}},\ }\href@noop {} {\bibfield  {journal}
  {\bibinfo  {journal} {Astrophys. J. Supp. Ser.}\ }\textbf {\bibinfo {volume}
  {216}},\ \bibinfo {pages} {15} (\bibinfo {year} {2015})}\BibitemShut
  {NoStop}%
\bibitem [{\citenamefont {Gordon}\ \emph {et~al.}(2022)\citenamefont {Gordon},
  \citenamefont {Rothman}, \citenamefont {Hargreaves}, \citenamefont {Hashemi},
  \citenamefont {Karlovets}, \citenamefont {Skinner}, \citenamefont {Conway},
  \citenamefont {Hill}, \citenamefont {Kochanov}, \citenamefont {Tan},
  \citenamefont {Wcis{\l}o}, \citenamefont {Finenko}, \citenamefont {Nelson},
  \citenamefont {Bernath}, \citenamefont {Birk}, \citenamefont {Boudon},
  \citenamefont {Campargue}, \citenamefont {Chance}, \citenamefont {Coustenis},
  \citenamefont {Drouin}, \citenamefont {Flaud}, \citenamefont {Gamache},
  \citenamefont {Hodges}, \citenamefont {Jacquemart}, \citenamefont {Mlawer},
  \citenamefont {Nikitin}, \citenamefont {Perevalov}, \citenamefont {Rotger},
  \citenamefont {Tennyson}, \citenamefont {Toon}, \citenamefont {Tran},
  \citenamefont {Tyuterev}, \citenamefont {Adkins}, \citenamefont {Baker},
  \citenamefont {Barbe}, \citenamefont {Can{\`{e}}}, \citenamefont
  {Cs{\'{a}}sz{\'{a}}r}, \citenamefont {Dudaryonok}, \citenamefont {Egorov},
  \citenamefont {Fleisher}, \citenamefont {Fleurbaey}, \citenamefont
  {Foltynowicz}, \citenamefont {Furtenbacher}, \citenamefont {Harrison},
  \citenamefont {Hartmann}, \citenamefont {Horneman}, \citenamefont {Huang},
  \citenamefont {Karman}, \citenamefont {Karns}, \citenamefont {Kassi},
  \citenamefont {Kleiner}, \citenamefont {Kofman}, \citenamefont
  {Kwabia-Tchana}, \citenamefont {Lavrentieva}, \citenamefont {Lee},
  \citenamefont {Long}, \citenamefont {Lukashevskaya}, \citenamefont {Lyulin},
  \citenamefont {Makhnev}, \citenamefont {Matt}, \citenamefont {Massie},
  \citenamefont {Melosso}, \citenamefont {Mikhailenko}, \citenamefont
  {Mondelain}, \citenamefont {M{\"{u}}ller}, \citenamefont {Naumenko},
  \citenamefont {Perrin}, \citenamefont {Polyansky}, \citenamefont {Raddaoui},
  \citenamefont {Raston}, \citenamefont {Reed}, \citenamefont {Rey},
  \citenamefont {Richard}, \citenamefont {T{\'{o}}bi{\'{a}}s}, \citenamefont
  {Sadiek}, \citenamefont {Schwenke}, \citenamefont {Starikova}, \citenamefont
  {Sung}, \citenamefont {Tamassia}, \citenamefont {Tashkun}, \citenamefont
  {Vander~Auwera}, \citenamefont {Vasilenko}, \citenamefont {Vigasin},
  \citenamefont {Villanueva}, \citenamefont {Vispoel}, \citenamefont {Wagner},
  \citenamefont {Yachmenev},\ and\ \citenamefont {Yurchenko}}]{HITRAN2020}%
  \BibitemOpen
  \bibfield  {author} {\bibinfo {author} {\bibfnamefont {I.~E.}\ \bibnamefont
  {Gordon}}, \bibinfo {author} {\bibfnamefont {L.~S.}\ \bibnamefont {Rothman}},
  \bibinfo {author} {\bibfnamefont {R.~J.}\ \bibnamefont {Hargreaves}},
  \bibinfo {author} {\bibfnamefont {R.}~\bibnamefont {Hashemi}}, \bibinfo
  {author} {\bibfnamefont {E.~V.}\ \bibnamefont {Karlovets}}, \bibinfo {author}
  {\bibfnamefont {F.~M.}\ \bibnamefont {Skinner}}, \bibinfo {author}
  {\bibfnamefont {E.~K.}\ \bibnamefont {Conway}}, \bibinfo {author}
  {\bibfnamefont {C.}~\bibnamefont {Hill}}, \bibinfo {author} {\bibfnamefont
  {R.~V.}\ \bibnamefont {Kochanov}}, \bibinfo {author} {\bibfnamefont
  {Y.}~\bibnamefont {Tan}}, \bibinfo {author} {\bibfnamefont {P.}~\bibnamefont
  {Wcis{\l}o}}, \bibinfo {author} {\bibfnamefont {A.~A.}\ \bibnamefont
  {Finenko}}, \bibinfo {author} {\bibfnamefont {K.}~\bibnamefont {Nelson}},
  \bibinfo {author} {\bibfnamefont {P.~F.}\ \bibnamefont {Bernath}}, \bibinfo
  {author} {\bibfnamefont {M.}~\bibnamefont {Birk}}, \bibinfo {author}
  {\bibfnamefont {V.}~\bibnamefont {Boudon}}, \bibinfo {author} {\bibfnamefont
  {A.}~\bibnamefont {Campargue}}, \bibinfo {author} {\bibfnamefont {K.~V.}\
  \bibnamefont {Chance}}, \bibinfo {author} {\bibfnamefont {A.}~\bibnamefont
  {Coustenis}}, \bibinfo {author} {\bibfnamefont {B.~J.}\ \bibnamefont
  {Drouin}}, \bibinfo {author} {\bibfnamefont {J.~M.}\ \bibnamefont {Flaud}},
  \bibinfo {author} {\bibfnamefont {R.~R.}\ \bibnamefont {Gamache}}, \bibinfo
  {author} {\bibfnamefont {J.~T.}\ \bibnamefont {Hodges}}, \bibinfo {author}
  {\bibfnamefont {D.}~\bibnamefont {Jacquemart}}, \bibinfo {author}
  {\bibfnamefont {E.~J.}\ \bibnamefont {Mlawer}}, \bibinfo {author}
  {\bibfnamefont {A.~V.}\ \bibnamefont {Nikitin}}, \bibinfo {author}
  {\bibfnamefont {V.~I.}\ \bibnamefont {Perevalov}}, \bibinfo {author}
  {\bibfnamefont {M.}~\bibnamefont {Rotger}}, \bibinfo {author} {\bibfnamefont
  {J.}~\bibnamefont {Tennyson}}, \bibinfo {author} {\bibfnamefont {G.~C.}\
  \bibnamefont {Toon}}, \bibinfo {author} {\bibfnamefont {H.}~\bibnamefont
  {Tran}}, \bibinfo {author} {\bibfnamefont {V.~G.}\ \bibnamefont {Tyuterev}},
  \bibinfo {author} {\bibfnamefont {E.~M.}\ \bibnamefont {Adkins}}, \bibinfo
  {author} {\bibfnamefont {A.}~\bibnamefont {Baker}}, \bibinfo {author}
  {\bibfnamefont {A.}~\bibnamefont {Barbe}}, \bibinfo {author} {\bibfnamefont
  {E.}~\bibnamefont {Can{\`{e}}}}, \bibinfo {author} {\bibfnamefont {A.~G.}\
  \bibnamefont {Cs{\'{a}}sz{\'{a}}r}}, \bibinfo {author} {\bibfnamefont
  {A.}~\bibnamefont {Dudaryonok}}, \bibinfo {author} {\bibfnamefont
  {O.}~\bibnamefont {Egorov}}, \bibinfo {author} {\bibfnamefont {A.~J.}\
  \bibnamefont {Fleisher}}, \bibinfo {author} {\bibfnamefont {H.}~\bibnamefont
  {Fleurbaey}}, \bibinfo {author} {\bibfnamefont {A.}~\bibnamefont
  {Foltynowicz}}, \bibinfo {author} {\bibfnamefont {T.}~\bibnamefont
  {Furtenbacher}}, \bibinfo {author} {\bibfnamefont {J.~J.}\ \bibnamefont
  {Harrison}}, \bibinfo {author} {\bibfnamefont {J.~M.}\ \bibnamefont
  {Hartmann}}, \bibinfo {author} {\bibfnamefont {V.~M.}\ \bibnamefont
  {Horneman}}, \bibinfo {author} {\bibfnamefont {X.}~\bibnamefont {Huang}},
  \bibinfo {author} {\bibfnamefont {T.}~\bibnamefont {Karman}}, \bibinfo
  {author} {\bibfnamefont {J.}~\bibnamefont {Karns}}, \bibinfo {author}
  {\bibfnamefont {S.}~\bibnamefont {Kassi}}, \bibinfo {author} {\bibfnamefont
  {I.}~\bibnamefont {Kleiner}}, \bibinfo {author} {\bibfnamefont
  {V.}~\bibnamefont {Kofman}}, \bibinfo {author} {\bibfnamefont
  {F.}~\bibnamefont {Kwabia-Tchana}}, \bibinfo {author} {\bibfnamefont {N.~N.}\
  \bibnamefont {Lavrentieva}}, \bibinfo {author} {\bibfnamefont {T.~J.}\
  \bibnamefont {Lee}}, \bibinfo {author} {\bibfnamefont {D.~A.}\ \bibnamefont
  {Long}}, \bibinfo {author} {\bibfnamefont {A.~A.}\ \bibnamefont
  {Lukashevskaya}}, \bibinfo {author} {\bibfnamefont {O.~M.}\ \bibnamefont
  {Lyulin}}, \bibinfo {author} {\bibfnamefont {V.~Y.}\ \bibnamefont {Makhnev}},
  \bibinfo {author} {\bibfnamefont {W.}~\bibnamefont {Matt}}, \bibinfo {author}
  {\bibfnamefont {S.~T.}\ \bibnamefont {Massie}}, \bibinfo {author}
  {\bibfnamefont {M.}~\bibnamefont {Melosso}}, \bibinfo {author} {\bibfnamefont
  {S.~N.}\ \bibnamefont {Mikhailenko}}, \bibinfo {author} {\bibfnamefont
  {D.}~\bibnamefont {Mondelain}}, \bibinfo {author} {\bibfnamefont {H.~S.~P.}\
  \bibnamefont {M{\"{u}}ller}}, \bibinfo {author} {\bibfnamefont {O.~V.}\
  \bibnamefont {Naumenko}}, \bibinfo {author} {\bibfnamefont {A.}~\bibnamefont
  {Perrin}}, \bibinfo {author} {\bibfnamefont {O.~L.}\ \bibnamefont
  {Polyansky}}, \bibinfo {author} {\bibfnamefont {E.}~\bibnamefont {Raddaoui}},
  \bibinfo {author} {\bibfnamefont {P.~L.}\ \bibnamefont {Raston}}, \bibinfo
  {author} {\bibfnamefont {Z.~D.}\ \bibnamefont {Reed}}, \bibinfo {author}
  {\bibfnamefont {M.}~\bibnamefont {Rey}}, \bibinfo {author} {\bibfnamefont
  {C.}~\bibnamefont {Richard}}, \bibinfo {author} {\bibfnamefont
  {R.}~\bibnamefont {T{\'{o}}bi{\'{a}}s}}, \bibinfo {author} {\bibfnamefont
  {I.}~\bibnamefont {Sadiek}}, \bibinfo {author} {\bibfnamefont {D.~W.}\
  \bibnamefont {Schwenke}}, \bibinfo {author} {\bibfnamefont {E.}~\bibnamefont
  {Starikova}}, \bibinfo {author} {\bibfnamefont {K.}~\bibnamefont {Sung}},
  \bibinfo {author} {\bibfnamefont {F.}~\bibnamefont {Tamassia}}, \bibinfo
  {author} {\bibfnamefont {S.~A.}\ \bibnamefont {Tashkun}}, \bibinfo {author}
  {\bibfnamefont {J.}~\bibnamefont {Vander~Auwera}}, \bibinfo {author}
  {\bibfnamefont {I.~A.}\ \bibnamefont {Vasilenko}}, \bibinfo {author}
  {\bibfnamefont {A.~A.}\ \bibnamefont {Vigasin}}, \bibinfo {author}
  {\bibfnamefont {G.~L.}\ \bibnamefont {Villanueva}}, \bibinfo {author}
  {\bibfnamefont {B.}~\bibnamefont {Vispoel}}, \bibinfo {author} {\bibfnamefont
  {G.}~\bibnamefont {Wagner}}, \bibinfo {author} {\bibfnamefont
  {A.}~\bibnamefont {Yachmenev}}, \ and\ \bibinfo {author} {\bibfnamefont
  {S.~N.}\ \bibnamefont {Yurchenko}},\ }\href {\doibase
  10.1016/j.jqsrt.2021.107949} {\bibfield  {journal} {\bibinfo  {journal} {J.
  Quant. Spectrosc. Radiat. Transf.}\ }\textbf {\bibinfo {volume} {277}},\
  \bibinfo {pages} {107949} (\bibinfo {year} {2022})}\BibitemShut {NoStop}%
\bibitem [{\citenamefont {Cygan}\ \emph
  {et~al.}(2016{\natexlab{a}})\citenamefont {Cygan}, \citenamefont
  {W{\'o}jtewicz}, \citenamefont {Kowzan}, \citenamefont {Zaborowski},
  \citenamefont {Wcis{\l}o}, \citenamefont {Nawrocki}, \citenamefont {Krehlik},
  \citenamefont {{\'S}liwczy{\'n}ski}, \citenamefont {Lipi{\'n}ski},
  \citenamefont {Mas{\l}owski}, \citenamefont {Ciury{\l}o},\ and\ \citenamefont
  {Lisak}}]{Cygan2016JCP}%
  \BibitemOpen
  \bibfield  {author} {\bibinfo {author} {\bibfnamefont {A.}~\bibnamefont
  {Cygan}}, \bibinfo {author} {\bibfnamefont {S.}~\bibnamefont
  {W{\'o}jtewicz}}, \bibinfo {author} {\bibfnamefont {G.}~\bibnamefont
  {Kowzan}}, \bibinfo {author} {\bibfnamefont {M.}~\bibnamefont {Zaborowski}},
  \bibinfo {author} {\bibfnamefont {P.}~\bibnamefont {Wcis{\l}o}}, \bibinfo
  {author} {\bibfnamefont {J.}~\bibnamefont {Nawrocki}}, \bibinfo {author}
  {\bibfnamefont {P.}~\bibnamefont {Krehlik}}, \bibinfo {author} {\bibfnamefont
  {{\L}.}~\bibnamefont {{\'S}liwczy{\'n}ski}}, \bibinfo {author} {\bibfnamefont
  {M.}~\bibnamefont {Lipi{\'n}ski}}, \bibinfo {author} {\bibfnamefont
  {P.}~\bibnamefont {Mas{\l}owski}}, \bibinfo {author} {\bibfnamefont
  {R.}~\bibnamefont {Ciury{\l}o}}, \ and\ \bibinfo {author} {\bibfnamefont
  {D.}~\bibnamefont {Lisak}},\ }\href@noop {} {\bibfield  {journal} {\bibinfo
  {journal} {J. Chem. Phys.}\ }\textbf {\bibinfo {volume} {144}} (\bibinfo
  {year} {2016}{\natexlab{a}})}\BibitemShut {NoStop}%
\bibitem [{\citenamefont {Cygan}\ \emph {et~al.}(2019)\citenamefont {Cygan},
  \citenamefont {Wcis{\l}o}, \citenamefont {W{\'o}jtewicz}, \citenamefont
  {Kowzan}, \citenamefont {Zaborowski}, \citenamefont {Charczun}, \citenamefont
  {Bielska}, \citenamefont {Trawi{\'n}ski}, \citenamefont {Ciury{\l}o},
  \citenamefont {Mas{\l}owski},\ and\ \citenamefont
  {Lisak}}]{Cygan2019OE-3method}%
  \BibitemOpen
  \bibfield  {author} {\bibinfo {author} {\bibfnamefont {A.}~\bibnamefont
  {Cygan}}, \bibinfo {author} {\bibfnamefont {P.}~\bibnamefont {Wcis{\l}o}},
  \bibinfo {author} {\bibfnamefont {S.}~\bibnamefont {W{\'o}jtewicz}}, \bibinfo
  {author} {\bibfnamefont {G.}~\bibnamefont {Kowzan}}, \bibinfo {author}
  {\bibfnamefont {M.}~\bibnamefont {Zaborowski}}, \bibinfo {author}
  {\bibfnamefont {D.}~\bibnamefont {Charczun}}, \bibinfo {author}
  {\bibfnamefont {K.}~\bibnamefont {Bielska}}, \bibinfo {author} {\bibfnamefont
  {R.~S.}\ \bibnamefont {Trawi{\'n}ski}}, \bibinfo {author} {\bibfnamefont
  {R.}~\bibnamefont {Ciury{\l}o}}, \bibinfo {author} {\bibfnamefont
  {P.}~\bibnamefont {Mas{\l}owski}}, \ and\ \bibinfo {author} {\bibfnamefont
  {D.}~\bibnamefont {Lisak}},\ }\href@noop {} {\bibfield  {journal} {\bibinfo
  {journal} {Opt. Expr.}\ }\textbf {\bibinfo {volume} {27}},\ \bibinfo {pages}
  {21810} (\bibinfo {year} {2019})}\BibitemShut {NoStop}%
\bibitem [{\citenamefont {Wang}\ \emph {et~al.}(2017)\citenamefont {Wang},
  \citenamefont {Sun}, \citenamefont {Tao}, \citenamefont {Liu},\ and\
  \citenamefont {Hu}}]{Wang2017JCP}%
  \BibitemOpen
  \bibfield  {author} {\bibinfo {author} {\bibfnamefont {J.}~\bibnamefont
  {Wang}}, \bibinfo {author} {\bibfnamefont {Y.~R.}\ \bibnamefont {Sun}},
  \bibinfo {author} {\bibfnamefont {L.~G.}\ \bibnamefont {Tao}}, \bibinfo
  {author} {\bibfnamefont {A.~W.}\ \bibnamefont {Liu}}, \ and\ \bibinfo
  {author} {\bibfnamefont {S.~M.}\ \bibnamefont {Hu}},\ }\href {\doibase
  10.1063/1.4998763} {\bibfield  {journal} {\bibinfo  {journal} {J. Chem.
  Phys.}\ }\textbf {\bibinfo {volume} {147}},\ \bibinfo {pages} {091103}
  (\bibinfo {year} {2017})}\BibitemShut {NoStop}%
\bibitem [{\citenamefont {Wang}\ \emph {et~al.}(2021)\citenamefont {Wang},
  \citenamefont {Hu}, \citenamefont {Liu}, \citenamefont {Sun}, \citenamefont
  {Tan},\ and\ \citenamefont {Hu}}]{Wang2021JQSRT-CO}%
  \BibitemOpen
  \bibfield  {author} {\bibinfo {author} {\bibfnamefont {J.}~\bibnamefont
  {Wang}}, \bibinfo {author} {\bibfnamefont {C.-L.}\ \bibnamefont {Hu}},
  \bibinfo {author} {\bibfnamefont {A.-W.}\ \bibnamefont {Liu}}, \bibinfo
  {author} {\bibfnamefont {Y.}~\bibnamefont {Sun}}, \bibinfo {author}
  {\bibfnamefont {Y.}~\bibnamefont {Tan}}, \ and\ \bibinfo {author}
  {\bibfnamefont {S.-M.}\ \bibnamefont {Hu}},\ }\href {\doibase
  10.1016/j.jqsrt.2021.107717} {\bibfield  {journal} {\bibinfo  {journal} {J.
  Quant. Spectrosc. Radiat. Transf.}\ }\textbf {\bibinfo {volume} {270}},\
  \bibinfo {pages} {107717} (\bibinfo {year} {2021})}\BibitemShut {NoStop}%
\bibitem [{\citenamefont {Reed}\ \emph {et~al.}(2017)\citenamefont {Reed},
  \citenamefont {Polyansky},\ and\ \citenamefont {Hodges}}]{Reed2017-CO}%
  \BibitemOpen
  \bibfield  {author} {\bibinfo {author} {\bibfnamefont {Z.}~\bibnamefont
  {Reed}}, \bibinfo {author} {\bibfnamefont {O.}~\bibnamefont {Polyansky}}, \
  and\ \bibinfo {author} {\bibfnamefont {J.}~\bibnamefont {Hodges}},\ }in\
  \href@noop {} {\emph {\bibinfo {booktitle} {2017 Conference on Lasers and
  Electro-Optics (CLEO)}}}\ (\bibinfo {organization} {IEEE},\ \bibinfo {year}
  {2017})\ pp.\ \bibinfo {pages} {1--2}\BibitemShut {NoStop}%
\bibitem [{\citenamefont {Ngo}\ \emph {et~al.}(2013)\citenamefont {Ngo},
  \citenamefont {Lisak}, \citenamefont {Tran},\ and\ \citenamefont
  {Hartmann}}]{Ngo2013JQSRT-HTP}%
  \BibitemOpen
  \bibfield  {author} {\bibinfo {author} {\bibfnamefont {N.}~\bibnamefont
  {Ngo}}, \bibinfo {author} {\bibfnamefont {D.}~\bibnamefont {Lisak}}, \bibinfo
  {author} {\bibfnamefont {H.}~\bibnamefont {Tran}}, \ and\ \bibinfo {author}
  {\bibfnamefont {J.-M.}\ \bibnamefont {Hartmann}},\ }\href@noop {} {\bibfield
  {journal} {\bibinfo  {journal} {J. Quant. Spectrosc. Radiat. Transf.}\
  }\textbf {\bibinfo {volume} {129}},\ \bibinfo {pages} {89} (\bibinfo {year}
  {2013})}\BibitemShut {NoStop}%
\bibitem [{\citenamefont {Tran}\ \emph {et~al.}(2014)\citenamefont {Tran},
  \citenamefont {Ngo},\ and\ \citenamefont {Hartmann}}]{Tran2014JQSRT-Err}%
  \BibitemOpen
  \bibfield  {author} {\bibinfo {author} {\bibfnamefont {H.}~\bibnamefont
  {Tran}}, \bibinfo {author} {\bibfnamefont {N.}~\bibnamefont {Ngo}}, \ and\
  \bibinfo {author} {\bibfnamefont {J.-M.}\ \bibnamefont {Hartmann}},\ }\href
  {\doibase 10.1016/j.jqsrt.2013.10.015} {\bibfield  {journal} {\bibinfo
  {journal} {J. Quant. Spectrosc. Radiat. Transf.}\ }\textbf {\bibinfo {volume}
  {134}},\ \bibinfo {pages} {104} (\bibinfo {year} {2014})}\BibitemShut
  {NoStop}%
\bibitem [{\citenamefont {Cheng}\ \emph {et~al.}(2015)\citenamefont {Cheng},
  \citenamefont {Wang}, \citenamefont {Sun}, \citenamefont {Tan}, \citenamefont
  {Kang},\ and\ \citenamefont {Hu}}]{cheng2015}%
  \BibitemOpen
  \bibfield  {author} {\bibinfo {author} {\bibfnamefont {C.~F.}\ \bibnamefont
  {Cheng}}, \bibinfo {author} {\bibfnamefont {J.}~\bibnamefont {Wang}},
  \bibinfo {author} {\bibfnamefont {Y.~R.}\ \bibnamefont {Sun}}, \bibinfo
  {author} {\bibfnamefont {Y.}~\bibnamefont {Tan}}, \bibinfo {author}
  {\bibfnamefont {P.}~\bibnamefont {Kang}}, \ and\ \bibinfo {author}
  {\bibfnamefont {S.~M.}\ \bibnamefont {Hu}},\ }\href {\doibase
  10.1088/0026-1394/52/5/S385} {\bibfield  {journal} {\bibinfo  {journal}
  {Metrologia}\ }\textbf {\bibinfo {volume} {52}},\ \bibinfo {pages} {S385}
  (\bibinfo {year} {2015})}\BibitemShut {NoStop}%
\bibitem [{\citenamefont {Tao}\ \emph {et~al.}(2018)\citenamefont {Tao},
  \citenamefont {Liu}, \citenamefont {Pachucki}, \citenamefont {Komasa},
  \citenamefont {Sun}, \citenamefont {Wang},\ and\ \citenamefont
  {Hu}}]{Tao2018}%
  \BibitemOpen
  \bibfield  {author} {\bibinfo {author} {\bibfnamefont {L.-G.}\ \bibnamefont
  {Tao}}, \bibinfo {author} {\bibfnamefont {A.-W.}\ \bibnamefont {Liu}},
  \bibinfo {author} {\bibfnamefont {K.}~\bibnamefont {Pachucki}}, \bibinfo
  {author} {\bibfnamefont {J.}~\bibnamefont {Komasa}}, \bibinfo {author}
  {\bibfnamefont {Y.~R.}\ \bibnamefont {Sun}}, \bibinfo {author} {\bibfnamefont
  {J.}~\bibnamefont {Wang}}, \ and\ \bibinfo {author} {\bibfnamefont {S.-M.}\
  \bibnamefont {Hu}},\ }\href {\doibase 10.1103/PhysRevLett.120.153001}
  {\bibfield  {journal} {\bibinfo  {journal} {Phys. Rev. Lett.}\ }\textbf
  {\bibinfo {volume} {120}},\ \bibinfo {pages} {153001} (\bibinfo {year}
  {2018})}\BibitemShut {NoStop}%
\bibitem [{\citenamefont {Rutkowski}\ \emph {et~al.}(2017)\citenamefont
  {Rutkowski}, \citenamefont {Johansson}, \citenamefont {Zhao}, \citenamefont
  {Hausmaninger}, \citenamefont {Khodabakhsh}, \citenamefont {Axner},\ and\
  \citenamefont {Foltynowicz}}]{Rutkowski2017OE-cavdisp}%
  \BibitemOpen
  \bibfield  {author} {\bibinfo {author} {\bibfnamefont {L.}~\bibnamefont
  {Rutkowski}}, \bibinfo {author} {\bibfnamefont {A.~C.}\ \bibnamefont
  {Johansson}}, \bibinfo {author} {\bibfnamefont {G.}~\bibnamefont {Zhao}},
  \bibinfo {author} {\bibfnamefont {T.}~\bibnamefont {Hausmaninger}}, \bibinfo
  {author} {\bibfnamefont {A.}~\bibnamefont {Khodabakhsh}}, \bibinfo {author}
  {\bibfnamefont {O.}~\bibnamefont {Axner}}, \ and\ \bibinfo {author}
  {\bibfnamefont {A.}~\bibnamefont {Foltynowicz}},\ }\href@noop {} {\bibfield
  {journal} {\bibinfo  {journal} {Opt. Expr.}\ }\textbf {\bibinfo {volume}
  {25}},\ \bibinfo {pages} {21711} (\bibinfo {year} {2017})}\BibitemShut
  {NoStop}%
\bibitem [{\citenamefont {Libbrecht}\ and\ \citenamefont
  {Libbrecht}(2006)}]{Libbrecht2006AJP-refractive}%
  \BibitemOpen
  \bibfield  {author} {\bibinfo {author} {\bibfnamefont {K.}~\bibnamefont
  {Libbrecht}}\ and\ \bibinfo {author} {\bibfnamefont {M.}~\bibnamefont
  {Libbrecht}},\ }\href@noop {} {\bibfield  {journal} {\bibinfo  {journal}
  {Amer. J. Phys.}\ }\textbf {\bibinfo {volume} {74}},\ \bibinfo {pages} {1055}
  (\bibinfo {year} {2006})}\BibitemShut {NoStop}%
\bibitem [{\citenamefont {Cygan}\ \emph {et~al.}(2015)\citenamefont {Cygan},
  \citenamefont {Wcis{\l}o}, \citenamefont {W{\'o}jtewicz}, \citenamefont
  {Mas{\l}owski}, \citenamefont {Hodges}, \citenamefont {Ciury{\l}o},\ and\
  \citenamefont {Lisak}}]{Cygan2015OE-1DCMDS}%
  \BibitemOpen
  \bibfield  {author} {\bibinfo {author} {\bibfnamefont {A.}~\bibnamefont
  {Cygan}}, \bibinfo {author} {\bibfnamefont {P.}~\bibnamefont {Wcis{\l}o}},
  \bibinfo {author} {\bibfnamefont {S.}~\bibnamefont {W{\'o}jtewicz}}, \bibinfo
  {author} {\bibfnamefont {P.}~\bibnamefont {Mas{\l}owski}}, \bibinfo {author}
  {\bibfnamefont {J.~T.}\ \bibnamefont {Hodges}}, \bibinfo {author}
  {\bibfnamefont {R.}~\bibnamefont {Ciury{\l}o}}, \ and\ \bibinfo {author}
  {\bibfnamefont {D.}~\bibnamefont {Lisak}},\ }\href@noop {} {\bibfield
  {journal} {\bibinfo  {journal} {Opt. Expr.}\ }\textbf {\bibinfo {volume}
  {23}},\ \bibinfo {pages} {14472} (\bibinfo {year} {2015})}\BibitemShut
  {NoStop}%
\bibitem [{\citenamefont {Gamache}\ \emph {et~al.}(2021)\citenamefont
  {Gamache}, \citenamefont {Vispoel}, \citenamefont {Rey}, \citenamefont
  {Nikitin}, \citenamefont {Tyuterev}, \citenamefont {Egorov}, \citenamefont
  {Gordon},\ and\ \citenamefont {Boudon}}]{Gamache2021JQSRT-Q}%
  \BibitemOpen
  \bibfield  {author} {\bibinfo {author} {\bibfnamefont {R.~R.}\ \bibnamefont
  {Gamache}}, \bibinfo {author} {\bibfnamefont {B.}~\bibnamefont {Vispoel}},
  \bibinfo {author} {\bibfnamefont {M.}~\bibnamefont {Rey}}, \bibinfo {author}
  {\bibfnamefont {A.}~\bibnamefont {Nikitin}}, \bibinfo {author} {\bibfnamefont
  {V.}~\bibnamefont {Tyuterev}}, \bibinfo {author} {\bibfnamefont
  {O.}~\bibnamefont {Egorov}}, \bibinfo {author} {\bibfnamefont {I.~E.}\
  \bibnamefont {Gordon}}, \ and\ \bibinfo {author} {\bibfnamefont
  {V.}~\bibnamefont {Boudon}},\ }\href {\doibase 10.1016/j.jqsrt.2021.107713}
  {\bibfield  {journal} {\bibinfo  {journal} {J. Quant. Spectrosc. Radiat.
  Transf.}\ }\textbf {\bibinfo {volume} {271}},\ \bibinfo {pages} {107713}
  (\bibinfo {year} {2021})}\BibitemShut {NoStop}%
\bibitem [{\citenamefont {Tran}\ \emph {et~al.}(2013)\citenamefont {Tran},
  \citenamefont {Ngo},\ and\ \citenamefont {Hartmann}}]{Tran2013JQSRT-HTP}%
  \BibitemOpen
  \bibfield  {author} {\bibinfo {author} {\bibfnamefont {H.}~\bibnamefont
  {Tran}}, \bibinfo {author} {\bibfnamefont {N.}~\bibnamefont {Ngo}}, \ and\
  \bibinfo {author} {\bibfnamefont {J.-M.}\ \bibnamefont {Hartmann}},\ }\href
  {\doibase 10.1016/j.jqsrt.2013.06.015} {\bibfield  {journal} {\bibinfo
  {journal} {J. Quant. Spectrosc. Radiat. Transf.}\ }\textbf {\bibinfo {volume}
  {129}},\ \bibinfo {pages} {199} (\bibinfo {year} {2013})}\BibitemShut
  {NoStop}%
\bibitem [{\citenamefont {Cygan}\ \emph
  {et~al.}(2016{\natexlab{b}})\citenamefont {Cygan}, \citenamefont {Wojtewicz},
  \citenamefont {Zaborowski}, \citenamefont {Wcis{\l}o}, \citenamefont {Guo},
  \citenamefont {Ciury{\l}o},\ and\ \citenamefont {Lisak}}]{Cygan2016MST}%
  \BibitemOpen
  \bibfield  {author} {\bibinfo {author} {\bibfnamefont {A.}~\bibnamefont
  {Cygan}}, \bibinfo {author} {\bibfnamefont {S.}~\bibnamefont {Wojtewicz}},
  \bibinfo {author} {\bibfnamefont {M.}~\bibnamefont {Zaborowski}}, \bibinfo
  {author} {\bibfnamefont {P.}~\bibnamefont {Wcis{\l}o}}, \bibinfo {author}
  {\bibfnamefont {R.}~\bibnamefont {Guo}}, \bibinfo {author} {\bibfnamefont
  {R.}~\bibnamefont {Ciury{\l}o}}, \ and\ \bibinfo {author} {\bibfnamefont
  {D.}~\bibnamefont {Lisak}},\ }\href@noop {} {\bibfield  {journal} {\bibinfo
  {journal} {Measurement Science and Technology}\ }\textbf {\bibinfo {volume}
  {27}},\ \bibinfo {pages} {045501} (\bibinfo {year}
  {2016}{\natexlab{b}})}\BibitemShut {NoStop}%
\bibitem [{\citenamefont {Nie}\ \emph {et~al.}(2024)\citenamefont {Nie},
  \citenamefont {Wang}, \citenamefont {Wang}, \citenamefont {Yang},
  \citenamefont {Hu}, \citenamefont {Sun},\ and\ \citenamefont {Hu}}]{Nie2024}%
  \BibitemOpen
  \bibfield  {author} {\bibinfo {author} {\bibfnamefont {Z.}~\bibnamefont
  {Nie}}, \bibinfo {author} {\bibfnamefont {X.}~\bibnamefont {Wang}}, \bibinfo
  {author} {\bibfnamefont {J.}~\bibnamefont {Wang}}, \bibinfo {author}
  {\bibfnamefont {Y.}~\bibnamefont {Yang}}, \bibinfo {author} {\bibfnamefont
  {C.}~\bibnamefont {Hu}}, \bibinfo {author} {\bibfnamefont {Y.}~\bibnamefont
  {Sun}}, \ and\ \bibinfo {author} {\bibfnamefont {S.}~\bibnamefont {Hu}},\
  }\href {https://link.cnki.net/urlid/62.1125.O4.20240219.1508.002} {\bibfield
  {journal} {\bibinfo  {journal} {Vaccum and Cryogenics}\ } (\bibinfo {year}
  {2024})}\BibitemShut {NoStop}%
\bibitem [{\citenamefont {Overney}\ \emph {et~al.}(2011)\citenamefont
  {Overney}, \citenamefont {Rufenacht}, \citenamefont {Braun}, \citenamefont
  {Jeanneret},\ and\ \citenamefont {Wright}}]{Overney2011IEEE}%
  \BibitemOpen
  \bibfield  {author} {\bibinfo {author} {\bibfnamefont {F.}~\bibnamefont
  {Overney}}, \bibinfo {author} {\bibfnamefont {A.}~\bibnamefont {Rufenacht}},
  \bibinfo {author} {\bibfnamefont {J.-P.}\ \bibnamefont {Braun}}, \bibinfo
  {author} {\bibfnamefont {B.}~\bibnamefont {Jeanneret}}, \ and\ \bibinfo
  {author} {\bibfnamefont {P.~S.}\ \bibnamefont {Wright}},\ }\href {\doibase
  10.1109/TIM.2011.2113950} {\bibfield  {journal} {\bibinfo  {journal} {IEEE
  Trans. Instrum. Meas.}\ }\textbf {\bibinfo {volume} {60}},\ \bibinfo {pages}
  {2172} (\bibinfo {year} {2011})}\BibitemShut {NoStop}%
\bibitem [{\citenamefont {W{\'o}jtewicz}\ \emph {et~al.}(2013)\citenamefont
  {W{\'o}jtewicz}, \citenamefont {Stec}, \citenamefont {Mas{\l}owski},
  \citenamefont {Cygan}, \citenamefont {Lisak}, \citenamefont {Trawi{\'n}ski},\
  and\ \citenamefont {Ciury{\l}o}}]{Wojtewicz2013JQSRT-CO-Profile}%
  \BibitemOpen
  \bibfield  {author} {\bibinfo {author} {\bibfnamefont {S.}~\bibnamefont
  {W{\'o}jtewicz}}, \bibinfo {author} {\bibfnamefont {K.}~\bibnamefont {Stec}},
  \bibinfo {author} {\bibfnamefont {P.}~\bibnamefont {Mas{\l}owski}}, \bibinfo
  {author} {\bibfnamefont {A.}~\bibnamefont {Cygan}}, \bibinfo {author}
  {\bibfnamefont {D.}~\bibnamefont {Lisak}}, \bibinfo {author} {\bibfnamefont
  {R.~S.}\ \bibnamefont {Trawi{\'n}ski}}, \ and\ \bibinfo {author}
  {\bibfnamefont {R.}~\bibnamefont {Ciury{\l}o}},\ }\href {\doibase
  10.1016/j.jqsrt.2013.06.005} {\bibfield  {journal} {\bibinfo  {journal} {J.
  Quant. Spectrosc. Radiat. Transf.}\ }\textbf {\bibinfo {volume} {130}},\
  \bibinfo {pages} {191} (\bibinfo {year} {2013})}\BibitemShut {NoStop}%
\end{thebibliography}
\providecommand{\noopsort}[1]{}\providecommand{\singleletter}[1]{#1}%

\end{document}